\pdfoutput=1

\documentclass[11pt,a4paper,oneside,english]{article}

\usepackage[english]{babel}
\usepackage[]{algorithm2e}

\usepackage{amsmath,amsfonts,amssymb,amsthm}
\usepackage{latexsym}
\usepackage{enumerate}
\usepackage[T1]{fontenc}
\usepackage[utf8x]{inputenc}
\usepackage{indentfirst}
\usepackage{fancyhdr}
\usepackage{graphicx}
\usepackage{bbm}
\usepackage{color}
\usepackage{mathrsfs}
\usepackage{hyperref}
\usepackage{nomencl}
\usepackage[textsize=small]{todonotes}

\theoremstyle{plain}

\theoremstyle{definition}

\theoremstyle{remark}

\numberwithin{equation}{section}

\renewcommand{\a}{\alpha}

\renewcommand{\o}{\omega}

\newcommand{\E}{\mathbb{E}}

\newcommand{\N}{\mathbb{N}}

\renewcommand{\P}{\mathbb{P}}

\newcommand{\R}{\mathbb{R}}
\newcommand{\T}{\mathbb{T}}

\newcommand{\CC}{\mathcal{C}}

\newcommand{\LL}{\mathcal{L}}
\newcommand{\MM}{\mathcal{M}}

\newcommand{\VV}{\mathcal{V}}
\newcommand{\WW}{\mathcal{W}}

\newcommand{\1}{\mathbbm{1}}

\newlength{\dhatheight}

\title{Rare event simulation for stochastic dynamics in continuous time} 

\author{Letizia Angeli, Stefan Grosskinsky, Adam M.\ Johansen,\\ Andrea Pizzoferrato} 

\date{\today} 

\begin{document}

\RestyleAlgo{boxruled}

\maketitle 

\abstract{
Large deviations for additive path functionals of stochastic dynamics and related numerical approaches have attracted significant recent research interest. We focus on the question of convergence properties for cloning algorithms in continuous time, and establish connections to the literature of particle filters and sequential Monte Carlo methods. This enables us to derive  rigorous convergence bounds for cloning algorithms which we report in this paper, with details of proofs given in a further publication. The tilted generator characterizing the large deviation rate function can be associated to non-linear processes which give rise to several representations of the dynamics and additional freedom for associated numerical approximations. We discuss these choices in detail, and combine insights from the filtering literature and cloning algorithms to compare different approaches and improve efficiency.
}

\tableofcontents

\section{Introduction\label{sec:intro}}
Large deviation simulation techniques based on classical ideas of evolutionary algorithms \cite{anderson,grassberger} have been proposed under the name of `cloning algorithms' in \cite{giardina} for discrete and in \cite{lecomte2007numerical} for continuous time processes, in order to study rare events of dynamic observables of interacting lattice gases. This approach has subsequently been applied in a wide variety of contexts (see e.g.\ \cite{giardina2011simulating,sollich,nemoto2016,hidalgo18} and references therein), and more recently, the convergence properties of the algorithm have become a subject of interest. Analytical approaches so far are based on a branching process interpretation of the algorithm in discrete time \cite{hidalgo1}, with limited and mostly numerical results in continuous time \cite{hidalgo2}. Systematic errors arise from the correlation structure of the cloning ensemble which can be large in practice, and several variants of the approach have been proposed to address those including e.g.\ a multicanonical feedback control \cite{nemoto2016}, adaptive sampling methods \cite{ferre} or systematic resampling \cite{brewer}. A recent survey of these issues and different variants of cloning algorithms in discrete and continuous time can be found in \cite{hurtado19}, Section III.

In this paper we provide a novel perspective on the underlying structure of the cloning algorithm, which is in fact well established in the statistics and applied probability literature on Feynman-Kac models and particle filters \cite{del2000moran,del2000branching,del2004feynman}. The framework we develop here can be used to generalize rigorous convergence results in \cite{rousset2006control} to the setting of continuous-time cloning algorithms as introduced in \cite{lecomte2007numerical}. Full mathematical details of this work are published in \cite{mathpaper}, and here we focus on describing the underlying approach and report the main convergence results. A second motivation is to use different McKean interpretations of Feynman-Kac semigroups (see Section \ref{sec:mckean}) to highlight several degrees of freedom in the design of cloning-type algorithms that can be used to improve performance. We illustrate this with the example of current large deviations for the inclusion process (originally introduced in \cite{giardina09}), aspects of which have previously been studied \cite{pizzo2}. Current fluctuations in stochastic lattice gases have attracted significant recent research interest (see e.g.\  \cite{lazarescu,hurtado2014,pizzo1} and references therein), and are one of the main application areas of cloning algorithms which are particularly challenging. 
In contrast to previous work in the context of cloning algorithms \cite{hidalgo1,hidalgo2}, our mathematical approach does not require a time discretization and works in a very general setting of a jump Markov process on a  compact state space. This covers in particular any finite state Markov chain or stochastic lattice gas on a finite lattice. 

The paper is organized as follows. In Section \ref{sec:setting} we introduce notation, the Feynman-Kac semigroup and several representations of the associated non-linear process. In Section \ref{sec:filter} we describe different particle approximations including the cloning algorithm, and summarize results published in \cite{mathpaper} on convergence properties of estimators based on the latter. In Section \ref{sec:ip} we describe a modification of the cloning algorithm for a particular class of stochastic lattice gases and apply it to the inclusion process as an example.

\section{Mathematical setting\label{sec:setting}}

\subsection{Large deviations and the tilted generator}

We consider a continuous-time Markov jump process $\big( X(t) :t\geq 0\big)$ on a compact state space $E$. To fix ideas we can think of a finite state Markov chain, such as a stochastic lattice gas on a finite lattice $\Lambda$ with a fixed number of particles $M$. Here $E$ is of the form $S^\Lambda$ with a finite set $S$ of local states (e.g. $S=\{ 0,1\}$ or $\{ 0,\ldots ,M\}$), but continuous settings with compact $E\subset \R^d$ for any $d\geq 1$ are also included. One can in principle also generalize to separable and locally compact state spaces, including countable Markov chains and lattice gases on finite lattices with open boundaries. But this would require more effort and complicate not only the proof but also the presentation of the main results for technical reasons which we want to avoid here (see \cite{mathpaper} for a more detailed discussion).

Jump rates are given by the kernel $W(x,dy)$, such that for all $x\in E$ and measurable subsets $A\subset E$
\begin{equation}\label{kernel}
\P \big[ X(t+\Delta t) \in A\big| X(t) =x\big] =\Delta t\int_A W(x,dy) +o(\Delta t)\quad\mbox{as }\Delta t\to 0\ ,
\end{equation}
where $o(\Delta t)/\Delta t\to 0$. 
We use the standard notation $\P$ and $\E$ for the distribution and the corresponding expectation on the usual path space for jump processes
\[
\Omega =\big\{ \omega :[0,\infty )\to E\mbox{ right continuous with left limits}\big\}\ .
\]
If we want to stress a particular initial condition $x\in E$ of the process we write $\P_x$ and $\E_x$. 
The process can be characterized by the infinitesimal generator
\begin{equation}\label{generator}
\LL f(x)=\int_E W(x,dy)[f(y)-f(x)], \quad \forall f\in \CC_b (E), \, x\in E, 
\end{equation}
acting on all continuous bounded functions $f\in \CC_b (E)$ on the state space. 
The adjoint $\LL^\dagger$ of this operator acts on probability distributions $\mu$ on $E$, and determines their time evolution via
\begin{equation}\label{master}
\frac{d}{dt} \mu_t (dy)=\int_{x\in E} \mu_t (dx) W(x,dy)-\int_{x\in E} \mu_t (dy) W(y,dx)\ ,
\end{equation}
where $\P [X_t \in A]=\int_A \mu_t (dy)$ for any regular $A\subset E$ characterizes the distribution of the process at time $t\geq 0$. In case of countable $E$, \eqref{master} is simply the usual master equation of the process for $\mu_t (y)=\P [X_t =y]$, but we focus our presentation on the equivalent description via the generator \eqref{generator}, which leads to a more compact notation and applies in the general setting. As a technical assumption we require that the total exit rate of the process is uniformly bounded
\begin{equation}\label{wbound}
w(x):=\int_E W(x,dy) \leq\bar w <\infty\quad\mbox{for all }x\in E\ .
\end{equation}

We are interested in the large deviations of additive path space observables $A_T:\Omega\to \R$ of the general form
\begin{equation}\label{oss_jump}
    A_T(\o):=\sum_{\substack{t\leq T\\\o(t_-)\neq \o(t)}}g\big(\o(t_-),\,\o(t)\big)\,+\,\int_0^T h\big(\o(t)\big)dt,
\end{equation}
where $g\in\CC_b (E^2)$ and $h\in\CC_b (E)$. Note that $A_T$ is well defined since the bound on $w (x)$ implies that the sum in the first term almost surely contains only finitely many terms for any $T>0$. The above functional, which recently appeared in this form in \cite{chetrite}, assigns a weight via the function $g$ to jumps of the process, as well as to the local time via the function $h$. Dynamics conditioned on such a functional have been studied in many contexts \cite{giardina2011simulating}, including driven diffusions on periodic continuous spaces $E$ \cite{nyawo}.

As mentioned before, the simplest examples covered by our setting are Markov chains with finite state space $E$. This includes stochastic particle systems on a finite lattice with periodic or closed boundary conditions such as zero-range or inclusion processes \cite{pizzo1,pizzo2,harris1}, and also processes with open boundaries and bounded local state space such as the exclusion process \cite{giardina}. Choosing $g$ appropriately and $h\equiv 0$ the functional $A_T$ can, for example, measure the empirical particle current across a bond of the lattice or within the whole system up to time $T$. 

We assume that $A_T$ admits a large deviation rate function, which is a lower semi-continuous function $I:\R\to [0,\infty ]$ such that
\begin{equation}
\lim_{T\to\infty} -\frac{1}{T}\log\P [A_T /T\in U] =\inf_{a\in U} I(a)
\end{equation}
for all regular intervals $U\subset\R$ (see e.g.\ \cite{hollander,dembo} for a more general discussion). Based on the graphical construction of the jump  process and the contraction principle, existence and convexity of $I$ can be established in a very general setting for countable state space (see e.g.\ \cite{bertini2015} and references therein). If $I$ is convex, it is characterized by the scaled cumulant generating function (SCGF)
\begin{equation}\label{scgf}
\lambda_k :=\lim_{T\to\infty} \frac{1}{T}\log \E \big[ e^{kA_T}\big]
\end{equation}
via the Legendre transform
\begin{equation}
I(a)=\sup_{k\in\R} \big( ka -\lambda_k \big)\ .
\end{equation}
It is well known (see e.g.\ \cite{harris1,chetrite}) that $\lambda_k$ can be characterized as the principal eigenvalue of a tilted version of the generator \eqref{generator}
\begin{align}\label{tilted}
\LL_k f (x)&:=\int_E W(x,dy)\big[ e^{kg(x,y)} f(y)-f(x)\big] +kh(x)f(x)\nonumber\\
&=\underbrace{\int_E W_k (x,dy)\big[ f(y)-f(x)\big]}_{=:\widehat{\LL}_k f(x)} +\VV_k (x)f(x)
\end{align}
with modified rates for the jump part $\hat\LL_k$
\begin{equation}\label{modrates}
W_k (x,dy):=W(x,dy) e^{kg(x,y)} \ ,
\end{equation}
and potential for the diagonal part
\begin{equation}\label{vk}
\VV_k (x):=\int_E W(x,dy)\big[ e^{kg(x,y)} -1\big] +kh(x)\ .
\end{equation}
By \eqref{wbound} and the boundedness of $g$ and $h$, for each $k\in\R$ there exist constants such that we have uniform bounds
\begin{equation}\label{vbound}
\int_E W_k (x,dy)\leq \bar w_k\quad\mbox{and}\quad \VV_k (x)\leq \bar v_k\quad\mbox{for all }x\in E\ .
\end{equation}
Note also that $\LL_0 =\LL$, but for any $k\neq 0$ the diagonal part of the operator does not vanish and
\begin{equation}\label{llk1}
\LL_k 1(x)=\VV_k (x)\quad\mbox{for all }x\in E\ .
\end{equation}
Still, it generates a Feynman-Kac semigroup (see e.g.\ \cite{rousset2006control,chetrite} for details), defined as
\begin{equation}\label{semig}
P_t^k f(x)=\big( e^{t\LL_k}\big)f(x):=\E_x\big[ fe^{kA_t}\big]\ ,
\end{equation}
which is the unique solution to the backward equation
\begin{equation}\label{back}
\frac{d}{dt}P_t^k f=P_t^k (\LL_k f)\quad\mbox{with}\quad P_0^k f=f\ .
\end{equation}
Due to the diagonal part of $\LL_k$ this does not conserve probability, i.e.\ for the constant function $f\equiv 1$ we get
\begin{equation}\label{noncon}
P_t^k 1(x)=\E_x \big[ e^{kA_t}\big]\neq 1\quad\mbox{for all }k\neq 0,\ t>0\ .
\end{equation}
The associated logarithmic growth rate
\begin{equation}\label{lkp}
\lambda_k (t):=\frac{1}{t}\log P_t^k 1(x)
\end{equation}
provides a finite-time approximation of the SCGF $\lambda_k$, which depends on the initial condition $x\in E$. We require convergence of this approximation as $t\to\infty$ as an asymptotic stability property of the process, discussed in detail in \cite{mathpaper} and references therein. Under exponential mixing assumptions, which are mild in our contexts of interest, it can be shown that for some constant $C>0$
\begin{equation}\label{lkt}
\big|\lambda_k (t)-\lambda_k\big|\leq C/t \quad\mbox{as }t\to\infty\ .
\end{equation}
This is for example the case if $E$ is finite and the process necessarily has a spectral gap, as is the case for all finite state lattice gases mentioned earlier. 
Furthermore, exponential mixing implies that the modified finite-time approximation 
\begin{equation}\label{lkap}
\lambda_k (at,t):=\frac{1}{(1-a)t}\log \frac{P_t^k 1(x)}{P_{at}^k 1(x)} \quad\mbox{with }a\in (0,1)\ ,
\end{equation}
with a 'burn-in' time period of length $at$, significantly improves the convergence in \eqref{lkt} to
\begin{equation}\label{lkat}
\big|\lambda_k (at,t)-\lambda_k\big|\leq C\rho^{at}/t \quad\mbox{as }t\to\infty
\end{equation}
for some $\rho\in (0,1)$. This is of course routinely used in Monte-Carlo sampling where systems are allowed to relax towards stationarity before measuring. 
These intrinsic properties of the process and the related finite-time errors in the estimation of $\lambda_k$ are not the main subject of this paper. In the following we simply assume asymptotic stability and \eqref{lkt}, and focus on the efficient numerical estimation of $\lambda_k (t)$ for any given $t\geq 0$. $\lambda_k (at,t)$ can be treated completely analogously, which is discussed in more detail in \cite{mathpaper}.\\


\subsection{McKean interpretation of the Feynman-Kac semigroup\label{sec:mckean}}

As usual, for a given initial distribution $\nu_0^k$ on $E$ the semigroup \eqref{semig} determines a measure $\nu_t^k$ at times $t\geq 0$ on $E$, which can be characterized weakly through integrals of bounded test functions $f\in C_b (E)$ as
\begin{equation}\label{unno}
\nu_t^k (f):=\int_E f(x)\nu_t^k (dx)=\int_E P_t^k f(x)\nu_0^k (dx)\ .
\end{equation}
Here and in the following we use the common short notation $\nu_t^k (f)$ for the integral of the function $f$ under the measure $\nu_t^k$ to simplify notation. 
Note that we can write \eqref{lkp} as
\begin{equation}\label{lkp2}
    \lambda_k (t)=\frac{1}{t}\log \nu_t^k (1)\ ,
\end{equation}
with a more general initial condition $\nu_0^k$. But since $P_t^k$ does not conserve probability, $\nu_t^k$ is not a normalized probability measure and it is consequently impossible to sample from it.

With \eqref{noncon} $\nu_t^k (1)>0$ for all $t\geq 0$, and we can define normalized versions of the measures via
\begin{equation}
\mu_t^k (f):=\nu_t^k (f)/\nu_t^k (1)\ .
\end{equation}
Using \eqref{tilted}, \eqref{llk1} and \eqref{back} on can derive the evolution equation
\begin{align}\label{muvolution}
\frac{d}{dt}\mu_t^k (f)&=\frac{d}{dt}\frac{\nu_t^k (f)}{\nu_t^k (1)} = \frac{1}{\nu_t^k (1)}\, \nu_t^k (\LL_k f)-\frac{\nu_t^k (f)}{\nu_t^k (1)^2}\,\nu_t^k (\LL_k 1)\nonumber\\
&= \mu_t^k (\LL_k f)-\mu_t^k (f)\, \mu_t^k (\LL_k 1) \nonumber\\
&= \mu_t^k (\widehat{\LL}_k f)+\mu_t^k (\VV_k f)-\mu_{t}^k (f)\,\mu_t^k (\VV_k)
\end{align}
with initial condition $\mu_0^k =\nu_0^k$. 
It can be shown by similar direct computation of $\frac{d}{dt}\log \nu_t^k (1)$, using \eqref{llk1} and \eqref{lkp2} that
\begin{equation}\label{lambdat}
\lambda_k (t)=\frac{1}{t}\int_0^t \mu_s^k (\VV_k )ds\ .
\end{equation}
So the finite-time approximation of $\lambda_k$ is given by an ergodic average with respect to the distribution $\mu_t^k$, depending on the initial distribution $\mu_0^k$, with an obvious modification for \eqref{lkap}. The asymptotic stability of the original process implies that $\mu_t^k \to\mu_\infty^k$ converges to a unique stationary distribution $\mu_\infty^k$ on $E$, so that the SCGF \eqref{scgf} can be written as the stationary expectation of the potential
\[
\lambda_k =\mu_\infty^k (\VV_k )\ .
\]
Due to the non-linear nature of \eqref{muvolution}, $\mu_\infty^k$ is characterized by stationarity as the solution of the non-linear equation
\[
\mu_\infty^k (\widehat{\LL}_k f)=\mu_\infty^k (f)\,\mu_\infty^k (\VV_k)-\mu_\infty^k (\VV_k f)\quad\mbox{for all }f\in C_b (E)\ .
\]

Usually $\mu_\infty^k$ cannot be evaluated explicitly, but from \eqref{muvolution} it is possible to define a generic processes $\big( X_k (t):t\geq 0\big)$ with time-marginals $\mu_t^k$, and then use Monte Carlo sampling techniques. The first term of \eqref{muvolution} already corresponds to a jump process with generator $\widehat{\LL}_k$, and we have to rewrite the second non-linear part to be of the form of a generator. There is some freedom at this stage, and we report three common choices from the applied probability literature on particle approximations \cite{del2003particle,rousset2006control}, one of which corresponds to the approach in \cite{giardina,lecomte2007numerical}, and to the best of the authors' knowledge the other two have not been considered in the computational physics literature so far. 

For every probability distribution $\mu$ on $E$ we can write
\begin{equation}\label{sele}
    \mu(\VV_k f)-\mu(f)\,\mu(\VV_k)\,=\, \mu\big(\LL^-_{k,\mu,c} f+\LL^+_{k,\mu,c} f\big),
\end{equation}
where
\begin{equation}\label{kill}
    \LL^-_{k,\mu,c}f(x):=\big(\VV_k(x)-c\big)^-\,\int_E \big( f(y)-f(x) \big) \, \mu(dy)
\end{equation}
and
\begin{equation}\label{clone}
    \LL_{k,\mu,c}^+ f(x):=\int_E \big(\VV_k(y)-c\big)^+\big( f(y)-f(x) \big) \, \mu(dy),
\end{equation}
using the standard notation $a^+ =\max\{ 0,a\} $ and $a^-=\max\{0,-a\}$ for positive and negative part of $a\in\R$. We have the freedom to introduce an arbitrary constant $c\in\R$, possibly depending also on the measure $\mu$ (but not the state $x\in E$), since the left-hand side of \eqref{sele} is invariant under renormalization of the potential $\VV_k (x)\to \VV_k (x)-c$. The generators $\LL^-_{k,\mu,c}$ and $\LL^+_{k,\mu,c}$ describe jump processes on $E$ with rates depending on the probability measure $\mu$. 
$\VV_k (x)$ can be interpreted as a fitness potential for the process, and play exactly that role in the particle approximation of this process based on population dynamics, which is presented in Section \ref{sec:filter}. Generic choices are:
\begin{itemize}
\item $c=0$ is the default and simplest choice, but is usually not optimal as  discussed in Section \ref{sec:ip}.
\item $c=\mu_t^k (\VV_k )$ corresponding to the average potential: If the system in state $x$ is less fit than $c$ it jumps to state $y$ chosen from the distribution $\mu_t^k (dy)$ according to \eqref{kill}, and independently, the system jumps to states fitter than $c$ irrespective of its current state according to \eqref{clone}.
\item $c=\sup_{x\in E} \VV_k (x)$ or $\inf_{x\in E} \VV_k (x)$, so that $\LL_{k,\mu,c}^+(f)(x)\equiv 0$ or $\LL_{k,\mu,c}^- (f)(x)\equiv 0$, respectively, and only one of the two processes has to be implemented in a simulation.
\end{itemize}

Another representation of the non-linear part in \eqref{muvolution} is (see e.g.\ \cite{del2013mean}, Section 5.3.1)
\begin{equation}\label{combine}
\LL_{k,\mu}^\VV f(x):=\int_E \big(\VV_k(y)-\VV_k (x)\big)^+\big( f(y)-f(x) \big)\mu (dy)\ ,
\end{equation}
which is particularly interesting for implementing efficient selection dynamics 
as discussed in Section \ref{sec:ip}. 
Here every jump from this part of the generator strictly increases the fitness of the process, which is a stronger version of the previous idea where the process on average increased its fitness above level $c$. The rate depends on departure state $x$ and target state $y$, which is in general computationally more expensive to implement than rates in \eqref{kill} and \eqref{clone}, but can still be feasible due to simplifications in many concrete examples as demonstrated in Section \ref{sec:ip}. 
A further improvement of that idea is given by
\begin{align}
\LL_{k,\mu}^\VV f(x):=&\big(\VV_k(x)-\mu (\VV_k )\big)^- \nonumber\\
&\int_E \frac{\big(\VV_k(y)-\mu(\VV_k )\big)^+}{\mu\big( (\VV_k -\mu(\VV_k ))^+\big)}\big( f(y)-f(x) \big)\mu (dy)\ ,\label{remres}
\end{align}
which resembles a continuous-time version of selection processes which are known under the names of stochastic remainder sampling \cite{smc:methodology:Bak85} or residual sampling \cite{smc:methodology:KLW94} in discrete time. Here selection events change the process from states $x$ of less than average fitness $\mu (\VV_k )$ to states $y$ fitter than average, but we will see in Section \ref{sec:ip} that this variant is harder to implement than \eqref{combine} in our area of interest, and offers only limited extra gain on selection efficiency. 

In summary, the evolution equation \eqref{muvolution} for $\mu_t^k$ can be written as
\begin{equation}\label{weakevo}
\frac{d}{dt} \mu_t^k (f)=\mu_t^k (\widehat{\LL}_k f)+\mu_t (\LL_{k,\mu_t}^\VV f)
\end{equation}
where the first choice with \eqref{kill} and \eqref{clone} is included defining $\LL_{k,\mu}^\VV =\LL_{k,\mu,c}^- +\LL_{k,\mu,c}^+$ in that case. This defines a Markov process $\big( X_k (t) :t\geq 0\big)$ on the state space $E$ with generator
\begin{equation}\label{mckeangen}
\LL_{k,\mu_t^k} f(x):=\widehat{\LL}_k f(x) +\LL_{k,\mu_t^k }^\VV f(x)\ .
\end{equation}
The process is non-linear since the transition rates in the generator $\LL_{k,\mu_t^k}$ depend on the distribution $\mu_t^k$ of the process at time $t$, and in particular the process is also time-inhomogeneous. While the generator is still a linear operator acting on test functions $f$, the adjoint $\LL_{k,\mu_t^k}^\dagger$ is a non-linear operator acting on measures $\mu_t^k$, generating their time evolution via
\begin{equation}
\frac{d}{dt}\mu_t^k (dy) =\LL_{k,\mu_t^k}^\dagger \mu_t^k \ ,
\end{equation}
which is equivalent to \eqref{weakevo}. 
This microscopic mass transport description consistent with the macroscopic description provided by the Feynman-Kac semigroup $P_t^k$ is also called a McKean representation \cite{del2004feynman,del2013mean}. It is well know that particle approximations of different McKean representations can have very different properties. 
The first part is similar to the original dynamics with modified rates $W_k$ \eqref{modrates}, and the second non-linear part depending on the distribution $\mu_t^k$ arises from normalizing the measures $\nu_t^k$. Note that $\mu_t^k$ and therefore the finite-time approximation $\lambda_k (t)$ in \eqref{lambdat} are uniquely determined by \eqref{muvolution}, and thus independent of the different McKean representations, as are of course the limiting quantities $\mu_\infty^k$ and $\lambda_k$. 
Also, these interpretations do not make use of concepts from population dynamics such as branching, which will only come into play when using particle approximations of the measures $\mu^k_t$ as explained in the next section.

\section{Particle approximations and the cloning algorithm\label{sec:filter}}

The rates of the non-linear process $\big( X_k (t):t\geq 0\big)$ \eqref{mckeangen} depend on the distribution $\mu_t$, which is not known a-priori in the cases in which we are interested. The natural framework to sample such non-linear processes approximately is a particle approximation, see e.g.\ \cite{del2003particle}. Here an ensemble $\big( \underline{X}_k (t) :t\geq 0\big)$ of $N$ processes (also called particles or clones) $X^i_k (t)$, $i=1,\ldots ,N$ is run in parallel on the state space $E^N$, and $\mu_t^k$ is approximated by the empirical distribution $\mu^N (\underline{X}_k (t))$ of the realizations, where for any $\underline{x}\in E^N$ we define
\begin{equation}\label{emea}
\mu^N (\underline{x}) (dy):=\frac{1}{N} \sum_{i=1}^N \delta_{x_i} (dy)\quad\mbox{as a distribution on }E\ .
\end{equation}
Since $\mu^N (\underline{X}_k (t))$ is fully determined by the state of the ensemble at time $t$, the particle approximation is a standard (linear) Markov process on $E^N$. 
This leads to an estimator for the SCGF using \eqref{lambdat} given by
\begin{equation}\label{lambdatn}
\Lambda^N_{k} (t):=\frac{1}{t}\int_0^t \mu^N (\underline{X}_k (s)) (\VV_k )ds=\frac{1}{t}\int_0^t \frac{1}{N}\sum_{i=1}^N \VV_k (X_k^i (s))ds\ ,
\end{equation}
which is a random object depending on the realization of the particle approximation. The full dynamics can be set up in various different ways such that $\mu^N (\underline{X}_k (t)) \to\mu_t^k$ converges as $N\to\infty$ for any $t\geq 0$. A generic version, directly related to the above McKean representations has been studied in the applied probability literature in great detail \cite{del2003particle,rousset2006control}, providing quantitative control on error bounds for convergence. After describing this approach, we present a different approach known in the theoretical physics literature under the name of cloning algorithms \cite{giardina2011simulating,hurtado19}, which provides some computational advantages but lacks general rigorous error control so far \cite{hidalgo1,hidalgo2}. We will then set up a framework to identify common aspects of both approaches, which can be used to generalize existing convergence results to obtain rigorous error bounds for cloning algorithms as described in detail in \cite{mathpaper}, and to compare computational efficiency of both approaches.

\subsection{Basic particle approximations}

The most basic particle approximation is simply to run the McKean dynamics \eqref{mckeangen} in parallel on each of the particles, replacing the dependence on $\mu_t^k$ by $\mu^N (\underline{X}_k (t))$ in the jump rates. 
Mathematically, denoting by $\LL^N_k$ the generator of the full $N$ particle process $\big(\underline{X}_k (t):t\geq 0\big)$ acting on functions $F:E^N \to\R$, this corresponds to
\begin{equation}\label{filtergen}
\LL^N_k F(\underline{x}) :=\sum_{i=1}^N \LL_{k,\mu^N (\underline{x}) }^i F(\underline{x})\ .
\end{equation}
Here $\LL_{k,\mu^N (\underline{x}) }^i$ is equivalent to \eqref{mckeangen} acting on particle $i$ only, i.e.\ on the function $x^i \mapsto F(\underline{x})$ while $x^j$, $j\neq i$ remain fixed. 
The linear part $\widehat{\LL}_k$ of \eqref{mckeangen} does not depend on $\mu_t^k$ and follows the original dynamics for each particle, referred to as `mutation' events in the standard population dynamics interpretation. In this context, the non-linear parts \eqref{kill} and \eqref{clone} can be interpreted as `selection' events leading to mean-field interactions between the particles. Using the definition \eqref{emea} of the empirical measures, we can write for the part \eqref{kill}
\begin{align}\label{llk-}
\LL^{-,i}_{k,\mu^N (\underline{x}),c}F(\underline{x})&=\big(\VV_k(x_i )-c\big)^-\,\int_E \big( F(\underline{x}^{i,y} )-F(\underline{x}) \big) \, \mu^N (\underline{x})(dy )\nonumber\\
&=\big(\VV_k(x_i )-c\big)^-\frac{1}{N}\sum_{j=1}^N \big( F(\underline{x}^{i,x_j} )-F(\underline{x}) \big)
\end{align}
with notation $\underline{x}^{i,y} =(x_1 ,\ldots ,x_{i-1} ,y,x_{i+1} ,\ldots ,x_N )$. So with a rate depending on the fitness of particle $i$, it is `killed' and replaced by a copy of particle $j$ uniformly chosen from all particles. 
Analogously, we have for \eqref{clone}
\begin{equation}\label{llk+}
\LL^{+,i}_{k,\mu^N (\underline{x}),c}F(\underline{x})=\frac{1}{N}\sum_{j=1}^N \big(\VV_k(x_j )-c\big)^+ \big( F(\underline{x}^{i,x_j} )-F(\underline{x})\big)\ ,
\end{equation}
which leads to the same transition $\underline{x}\to \underline{x}^{i,x_j }$, but with a different interpretation. Each particle $j$ in the system reproduces independently with a rate depending on its fitness (cloning event), and its offspring replaces a uniformly chosen particle, which is equal to $i$ with probability $1/N$. 
The different nature of killing and cloning events becomes clearer when we write out the full generator \eqref{filtergen} and switch summation indices for the cloning part \eqref{llk+} in the second line,
\begin{align}\label{filter1}
\LL^N_{k} F(\underline{x}) =&\sum_{i=1}^N \int_E W_k (x_i ,dy)\big( F(\underline{x}^{i,y})-F(\underline{x})\big)\nonumber\\
&+\sum_{i=1}^N \big(\VV_k(x_i )-c\big)^+ \frac{1}{N}\sum_{j=1}^N\big( F(\underline{x}^{j,x_i} )-F(\underline{x})\big)\nonumber\\
&+\sum_{i=1}^N\big(\VV_k(x_i )-c\big)^-\frac{1}{N}\sum_{j=1}^N \big( F(\underline{x}^{i,x_j})-F(\underline{x} )\big)\ .
\end{align}
Analogously, the McKean representations \eqref{combine} and \eqref{remres} lead to basic $N$-particle systems with generators
\begin{align}\label{filter2}
\LL^N_{k} F(\underline{x}) =&\sum_{i=1}^N \int_E W_k (x_i ,dy)\big( F(\underline{x}^{i,y})-F(\underline{x})\big)\nonumber\\
&+\frac{1}{N}\sum_{i,j=1}^N\big(\VV_k(x_j )-\VV_k(x_i )\big)^+ \big( F(\underline{x}^{i,x_j})-F(\underline{x} )\big)
\end{align}
and
\begin{align}\label{filter3}
\LL^N_{k} &F(\underline{x}) =\sum_{i=1}^N \int_E W_k (x_i ,dy)\big( F(\underline{x}^{i,y})-F(\underline{x})\big)\nonumber\\
&+\frac{1}{N}\sum_{i,j=1}^N\big(\VV_k(x_i ){-}\mu (\VV_k )\big)^- \frac{\big(\VV_k(x_j )-\mu (\VV_k )\big)^+}{\mu \big( (\VV_k -\mu (\VV_k ))^+\big)}\big( F(\underline{x}^{i,x_j})-F(\underline{x} )\big)\ .
\end{align}
Here a particle $i$ is replaced by a particle $j$ with higher fitness, 
combining killing and cloning into a single event. In the case of \eqref{filter3}, particle $i$ is furthermore less fit and $j$ is fitter than average. Note that these approximating systems  
can be seen as particle systems with mean-field or averaged pairwise interaction given by the selection dynamics.

Following established results in  \cite{del2003particle,rousset2006control,del2013mean}, 
the (random) quantity $\Lambda_k^N (t)$ is an asymptotically unbiased estimator of $\lambda_k (t)$ with a systematic error bounded by
\begin{equation}\label{bias}
\sup_{t\geq 0} \Big|\E^N \big[ \Lambda_{k}^N (t) \big] -\lambda_k (t)\Big| \leq \frac{C}{N}\quad\mbox{for all }N\geq 1\ ,
\end{equation}
along with several rigorous convergence results. These include an estimate on the random error in $L^p$ norm for any $p>1$,
\begin{equation}\label{lpbound}
\sup_{t\geq 0} \E^N \big[ |\Lambda_{k}^N (t) -\lambda_k (t)|^p \big]^{1/p} \leq \frac{C_p}{\sqrt{N}}\quad\mbox{for all }N\geq 1\ ,
\end{equation}
as well as other formulations  
including almost sure convergence. 
Note that these estimates are uniform in $t\geq 0$, so are not affected by the choice of simulation time. The use of a finite simulation time, $t$, leads to an additional systematic error to the estimate of the SCGF $\lambda_k$, of order $1/t$ as in \eqref{lkt} or $\rho^{at} /t$ as in \eqref{lkat}.  
The bound \eqref{lpbound} for $p=2$ implies for the variance
\begin{equation}\label{avar}
\sup_{t\geq 0} \E^N \Big[ \Big|\Lambda_{k}^N (t) -\E^N \big[ \Lambda_{k}^N (t) \big]\Big|^2 \Big] \leq \frac{C_2^2}{N}\quad\mbox{for all }N\geq 1\ ,
\end{equation}
since we have $\mathrm{Var}(Y)= \inf_{a\in\R} \E \big[ (Y-a)^2 \big]$ for any real-valued random variable $Y$. Therefore, error bars based on standard deviations are of the usual Monte Carlo order of $1/\sqrt{N}$, and the random error dominates the systematic bias \eqref{bias} for $N$ large enough. Further remarks on possible unbiased estimators can be found at the end of the next subsection.\\

\subsection{Essential properties of particle approximations\label{sec:essential}}

Following the standard martingale characterization of Feller-type Markov processes (see e.g.\ \cite{liggett2010continuous}, Chapter 3), we know that for every bounded, continuous $F\in C_b (E^N )$
\begin{equation}\label{marti}
\MM_F^N (t):=F\big(\underline{X}_k (t)\big) -F\big(\underline{X}_k (0)\big) -\int_0^t \LL_k^N F\big(\underline{X}_k (s)\big) ds
\end{equation}
is a martingale on $\R$ with (predictable) quadratic variation
\begin{equation}\label{qvar}
\langle\MM_F^N \rangle (t)=\int_0^t \Gamma_k^N F\big(\underline{X}_k (s)\big) ds\ ,
\end{equation}
where the associated carr\'e du champ operator $\Gamma_k^N$ is given by
\begin{equation}\label{cdo}
\Gamma_k^N F(\underline{x}) := \LL_k^N F^2 (\underline{x}) -2F (\underline{x})\LL_k^N F (\underline{x})\ .
\end{equation}
In analogy to the decomposition of a random variable into its mean and a centred fluctuating part, the martingale \eqref{marti} describes the fluctuations of the process $t\mapsto F\big(\underline{X}_k (t)\big)$. The strength of the noise depends on time and is given by the increasing process \eqref{qvar}, whose time evolution is generated by the carr\'e du champ operator \eqref{cdo}. In contrast to the generator $\LL_k^N$, this is a quadratic (non-linear) operator and it is the main tool for studying the fluctuations of a process.

Elementary computations for approximations \eqref{filter1}, \eqref{filter2} and \eqref{filter3} show that for marginal test functions $F (\underline{x})=f(x_i )$ depending only on a single particle, we have 
\[
\LL_k^N F (\underline{x}) =\LL_{k,\mu^N (\underline{x})} f(x_i )\quad\mbox{and}\quad \Gamma_k^N F (\underline{x}) =\Gamma_{k,\mu^N (\underline{x})} f(x_i )\ .
\]

So generator and carr\'e du champ both coincide with the corresponding operators $\LL_{k,\mu^N (\underline{x})}$ and $\Gamma_{k,\mu^N (\underline{x})}$ for the McKean dynamics \eqref{mckeangen}. 
This means that for large enough $N$ and $\mu^N \big(\underline{X}_k (t)\big)$ close to $\mu_t^k$, each marginal process $t\mapsto X_k^i (t)$ has essentially the same distribution as the corresponding McKean process $t\mapsto X_k (t)$. Note that due to selection events these (auxiliary) dynamics do not coincide with the original process conditioned on a large deviation event, and they are also not unique since there are various choices for McKean representations of Feynman-Kac semigroups, as discussed earlier. Trajectories in a particle approximation  always correspond to the trajectories of the particular McKean interpretation they are based on, which is usually \eqref{sele} 
in the context of cloning algorithms. Due to asymptotic stability the particle approximation converges as $t\to\infty$ for fixed $N$ to a unique stationary distribution $\mu_\infty^{N,k}$, and the single-particle marginals of this distribution converge to $\mu_\infty^k$ as $N\to\infty$. 
While the marginal processes for a given particle approximation are identically distributed they are not independent, and $\mu_\infty^{N,k}$ exhibits non-trivial correlations between particles resulting from selection events, which we discuss again in Section \ref{sec:ip}.

Now consider averaged observables of the form
\[
F(\underline{x}) =\frac{1}{N}\sum_{i=1}^N f(x_i )=\mu^N (\underline{x})(f)
\]
as they appear in the eigenvalue estimator \eqref{lambdatn}. 
Since the generator $\LL_k^N$ is a linear operator in $F$, we have the same identity as above for the generator,
\begin{equation}\label{avf}
\LL_k^N \mu^N (\underline{x})(f)=\mu^N (\underline{x})\big( \LL_{k,\mu^N (\underline{x})} f\big)\ .
\end{equation}
The carr\'e du champ, on the other hand, is non-linear in $F$ and the dependence between particles is captured by this operator. Since for all approximations \eqref{filter1}, \eqref{filter2} and \eqref{filter3} in every selection event only a single particle is affected, another elementary, slightly more cumbersome computation shows (see \cite{mathpaper} for details) 
\begin{equation}\label{avfgamma}
\Gamma_k^N \mu^N (\underline{x})(f) =\frac{1}{N} \mu^N (\underline{x})\big( \Gamma_{k,\mu^N (\underline{x})} f\big)\ .
\end{equation}
The factor $1/N$ results from a self-averaging property of the  mean-field interaction through selection dynamics, which is expected from results on other mean-field particle systems (see e.g.\ \cite{mim,daipra} and references therein), and is fully analogous to the central-limit type scaling of the empirical variance for the sum of $N$ independent random variables. 
While this scaling remains the same for more general particle approximations with more than one particle being affected by selection events, the simple identity \eqref{avfgamma} does not hold exactly for any $N\geq 1$ as we see in the next subsection.

Recall that the estimator \eqref{lambdatn} for the principal eigenvalue \eqref{scgf} is given by an ergodic integral of the average observable $F(\underline{x})=\mu^N (\underline{x})(\VV_k )$. With \eqref{vbound} $\VV_k \in C_b (E)$ and rates are bounded, so $\mu^N (\underline{x})\big( \Gamma_{k,\mu^N (\underline{x})} \VV_k \big)$ is also bounded and the carr\'e du champ \eqref{avfgamma} vanishes as $N\to\infty$. Therefore the martingale $\MM_F^N (t)$ also vanishes\footnote{in $L^p$-sense for any $p>1$ following with the Burkholder-Davis-Gundy inequality, see e.g.\ \cite{teicher}, Section 11} for all $t\geq 0$, leading to a convergence of the measures $\mu^N (\underline{X}_k (t))\to\mu_t^k (t)$ and also of finite time approximations $\Lambda_k^N (t)\to \lambda_k (t)$ as reported in the previous subsection. Due to the time-normalization in \eqref{lambdatn} and the assumed ergodicity, corresponding error bounds hold uniformly in $t\geq 0$. In summary, bounds on the carr\'e du champ are the main ingredient for the proof of convergence results as explained in detail in \cite{mathpaper} and references therein. All above properties up to and including \eqref{avf} are generic requirements for any particle approximation. These particle approximations can differ in their correlation structures and this freedom can be used to construct numerically more efficient particle approximations as discussed in the next subsection. To optimize sampling, particles should ideally evolve in as uncorrelated a fashion as possible; it is not possible to achieve completely independent evolution due to the non-linearity of the underlying McKean process and resulting selection events and mean-field interactions.\\

\noindent\textbf{Remarks on unbiased estimators.} Estimators based on expectations w.r.t.\ the empirical measures $\mu_t^N =\mu^N (\underline{X}_k (t))$ usually have a bias, i.e. $\E \big[ \mu_t^N (f)\big] \neq\mu_t (f)$ for $f\in\CC_b (E)$, which vanishes only asymptotically \eqref{bias}. This originates from the non-linear time evolution of $\mu_t^k$ \eqref{muvolution} and associated McKean processes. It is straightforward to derive estimators based on the unnormalized measures $\nu_t^k$ \eqref{unno} that are unbiased. Based on \eqref{lambdat} and \eqref{lambdatn}, we obtain an estimate of the normalization $\nu_t^k (1)$:
\begin{equation}\label{norma}
\nu_t^N (1) :=\exp \Big( \int_0^t \mu_s^N  (\VV_k )\Big)\ ,
\end{equation}
and then introduce unnormalized empirical measures on $E$ at time $t$ based on the particle approximation
\[
\nu_t^N (f):=\nu_t^N (1)\mu_t^N (f)\quad\mbox{for all }f\in\CC_b (E)\ .
\]
The expected time evolution of observables $f$ is then given by
\begin{equation}\label{vte}
\frac{d}{dt}\E \big[ \nu_t^N (f)\big] =\E\Big[ \nu_t^N (f)\mu_t^N (\VV_k )+\nu_t^N (1)\LL_k^N \mu_t^N (f)\Big]\ .
\end{equation}
Now with \eqref{avf} and the decomposition \eqref{mckeangen} of $\LL_{k,\mu_t^N}$ into mutation and selection part, we have
\[
\LL_k^N \mu_t^N (f)=\mu_t^N \big( \widehat{\LL}_k f+\LL_{k,\mu_t^N}^\VV f\big)\ ,
\]
and with the general construction of McKean representations \eqref{sele}
\[
\mu_t^N (\LL_{k,\mu_t^N}^\VV f)=\mu_t^N (\VV_k f)-\mu_t^N (f)\mu_t^N (\VV_k )\ .
\]
Inserting into \eqref{vte}, this simplifies to
\[
\frac{d}{dt}\E \big[ \nu_t^N (f)\big] =\E\big[ \nu_t^N (\widehat{\LL}_k f)+\nu_t^N (\VV_k f)\big]\ .
\]
Since with \eqref{back} $\LL_k =\widehat{\LL}_k +\VV_k$ also generates the time evolution of $\nu_t (f)$, a simple Gronwall argument with $\E \big[ \nu_0^N (f)\big] =\nu_0 (f)$ gives
\begin{equation}\label{gronwall1}
\E \big[ \nu_t^N (f)\big] =\nu_t (f)\quad\mbox{for all }t\geq 0 \mbox{ and }N\geq 1\ .
\end{equation}
Note that choosing $f\equiv 1$ implies that the normalization \eqref{norma} is an unbiased estimator of $e^{t\lambda_k (t)}$, which we will see again in Section \ref{sec:cfactor} in the context of cloning algorithms. However, in practice the random error dominates the accuracy of  estimates of $\lambda_k (t)$, so $N$ has to be chosen large and the bias of the estimator $\Lambda_k^N (t)$ \eqref{lambdatn} is negligible.

\subsection{Cloning algorithms\label{sec:cloning}}

Cloning algorithms proposed in the theoretical physics literature \cite{giardina,lecomte2007numerical} are similar to the particle approximation \eqref{filter1}, using the same tilted rates $W_k$ for mutations, but combining the cloning and mutation part of the generator. We focus the following exposition around the algorithm proposed in \cite{lecomte2007numerical}, but other continuous-time versions can be analysed analogously. The idea is simply to sample the cloning process for each particle $i$ together with the mutation process at the same rate
\[
w_k (x_i ):=\int_E W_k (x_i ,dy) =\int_E w(x_i ,dy)e^{kg(x_i ,y)}\ .
\]
In each combined event, a random number of clones is generated with a distribution $p_{k,x_i }^N (n)$ such that its expectation is
\begin{equation}\label{pmean}
m_k^N (x_i ):=\sum_{n=0}^N np_{k,x_i }^N (n)=\big(\VV_k (x_i )-c\big)^+ /w_k (x_i )\ .
\end{equation}
These clones then replace $n$ particles chosen uniformly at random (in the sense that all subsets of size $n$ are equally probable) from the ensemble.  
In this way, the rate at which a clone of particle $i$ replaces any given particle $j$ is
\[
w_k (x_i )\frac{m_k^N (x_i )}{N}=\frac{1}{N} \big(\VV_k (x_i )-c\big)^+ \ ,
\]
as required for $\LL_k^N$ in \eqref{filter1}. 
The only additional assumption on $p_{k,x_i }^N (n)$ is that the range of possible values for $n$ has to be bounded by $N$ for the cloning event to be well defined. Since its mean is bounded by $\max_{x\in E}\big(\VV_k (x)-c\big)^+ /w_k (x)<\infty$ independently of $N$, any distribution with the correct mean and finite range will lead to a valid algorithm for sufficiently large $N$.

The above cloning process is described by the generator
\begin{align}\label{mcgen}
\LL_k^{N,mc} F(\underline{x}):=\sum_{i=1}^N \sum_{n=0}^N\frac{1}{{N\choose n}}\sum_{A\subseteq\{ 1,..,N\}\atop |A|=n} \int_E & W_k (x_i ,dy) p_{k,x_i }^N (n)\nonumber\\
& \big( F(\underline{x}^{A,x_i ;i,y})-F(\underline{x})\big)\ .
\end{align}
Here we have used the notation $\underline{x}^{A,w;\,i,y}$ for the vector $\underline{z}\in E^N$ with
\begin{align*}
z_j:=
    \begin{cases}
    x_j & j\not\in A\cup\{i\}\\
    w & j\in A\setminus\{i\}\\
    y &j=i,
    \end{cases}
\end{align*}
for $j\in\{1,\dots,N\}$ and $w,y\in E$. \eqref{mcgen} combines cloning of $x_i$ into a uniformly chosen subset $A$ of size $n$, with a subsequent mutation event where the state of particle $i$ changes to $y$. If we simply write $\LL_k^{N,-}$ for the killing part (third line in \eqref{filter1}) which remains unchanged, the full generator of this cloning algorithm is given by
\begin{equation}\label{clonegen}
\LL_k^{N,clone}:= \LL_k^{N,mc} +\LL_k^{N,-}\ .
\end{equation}
It can be shown by direct computation that for marginal test functions of the form $F(\underline{x}) =f (x_i )$ 
this is equivalent to the generator \eqref{filter1}
\begin{equation}\label{margid}
\LL_k^N F(\underline{x})=\LL_k^{N,clone} F(\underline{x})\ ,
\end{equation}
and by linearity of generators also for all averaged functions of the form $F(\underline{x}) =\mu^N (\underline{x})(f)$. One can also show that for marginal test functions the carr\'e du champ operators coincide, so the cloning algorithm produces marginal processes or particle trajectories with the same distribution as the simple particle approximation \eqref{filter1}. For averaged test functions the change in the correlation structure between particles is picked up by the carr\'e du champ operator. Instead of \eqref{avfgamma} one can derive the following estimate for the mutation and cloning part
\[
\Gamma_k^{N,mc} F(\underline{x})\leq 
\frac{2}{N} \mu^N (\underline{x})\Big( \widehat{\Gamma}_k f+\1 (\VV_k >c)\frac{q_k^N }{m_k^N}\Gamma_{k,\mu^N (\underline{x}),c}^+ f\Big)\ ,
\]
see \cite{mathpaper}, Theorem 3.2, for a proof.

Here $q_k^N(x_i):=\sum_{n=0}^N n^2 p^N_{k,x_i}(n)$ denotes the second moment of the number of clones for the particle $i$, and we use the decomposition \eqref{mckeangen} where $\widehat{\Gamma}_k f$ is the carr\'e du champ corresponding to the mutation dynamics $\widehat{\LL}_k$, and $\Gamma_{k,\mu^N (\underline{x}),c}^+$ the one corresponding to the cloning part \eqref{clone}. 
This estimate holds, of course, only for $N$ large enough that the cloning event is well defined (see discussion above). 
Note also that $\Gamma_{k,\mu^N (\underline{x}),c}^+ f(x)$ is proportional to $(\VV_k (x)-c)^+$ and with \eqref{pmean} $(\VV_k (x)-c)^+ =0$ implies $m_k^N (x)=0$ for the expectation of the distribution $p_{k,x}^N$, leading to the indicator function $\1 (\VV_k >c) \in\{0,1\}$. 

This is sufficient to carry out the full proof of the convergence results mentioned in Section \ref{sec:filter} based on results in \cite{rousset2006control}. This is carried out in \cite{mathpaper} in full detail, and here we only report the main result of that work. Recall the bounds \eqref{vbound} on $\VV_k$ and the total modified exit rate $w_k$.\\

\noindent\textbf{Theorem.}{\it\ Denote by $\bar\Lambda_{k}^N (t)$ the eigenvalue estimator \eqref{lambdatn} corresponding to the cloning algorithm \eqref{clonegen}. Then there exist constants $\a,\,\gamma>0$ and $\a_p,\,\gamma_p >0$ such that for all $N$ large enough
\begin{equation}\label{th1}
\sup_{t\geq 0} \Big|\E^N \big[ \bar\Lambda_{k}^N (t) \big] -\lambda_k (t)\Big| \leq \frac{\a}{N}\bar v_k \bar w_k \big( \gamma+\| q_k^N \|_\infty \big)\ ,
\end{equation}
and for all $p>1$
\begin{equation}\label{th2}
\sup_{t\geq 0} \E^N \big[ |\bar\Lambda_{k}^N (t) -\lambda_k (t)|^p \big]^{1/p} \leq \frac{\a_p}{\sqrt{N}}\bar v_k \bar w_k \big( \gamma_p+\| q_k^N \|_\infty \big)^{1/2}\ .
\end{equation}
}
\medskip

\noindent\textbf{Remarks.}
\begin{itemize}
\item Choosing the normalization of the potential $c<\inf_{x\in E} \VV_k (x)$ the killing rate in \eqref{filter1} vanishes and \eqref{mcgen} describes the full generator $\LL_k^{N,clone}$ for the cloning algorithm. This is computationally cheaper and simpler to implement, since only the mutation process has to be sampled independently for all particles, and cloning events happen simultaneously. However, as is discussed in Section \ref{sec:ip}, this choice in general reduces the accuracy of the estimator.
\item A common choice in the physics literature for the distribution $p_{k,x_i}^N$ of the clone size event (see e.g. \cite{giardina,hurtado19}) is
  \begin{equation}\label{pchoice}
    p_{k,x_i}^N (n)=\left\{\begin{array}{cl} m_k^N (x_i ) -\lfloor m_k^N (x_i )\rfloor, &\mbox{ for }n=\lfloor m_k^N (x_i )\rfloor + 1\\ \lfloor m_k^N (x_i )\rfloor + 1 -m_k^N (x_i ), &\mbox{ for }n=\lfloor m_k^N (x_i )\rfloor\\ 0,&\mbox{ otherwise}\end{array}\right.
  \end{equation}
So the two adjacent integers to the mean are chosen with appropriate probabilities, which minimizes the variance of the distribution for a given mean. This choice therefore minimizes the contribution of the second moment $q_k^N$ to the bound for the errors in \eqref{th1} and \eqref{th2}, and is also simple to implement in practice.
\item Due to \eqref{margid}, trajectories of individual particles follow the same law as the simple particle approximation \eqref{filter1} and therefore the same McKean process as explained in Section \ref{sec:mckean} The cloning approach can introduce additional correlations between particles due to large cloning events, which is quantified by the second moment $q_k^N$ entering the error bounds in \eqref{th1} and \eqref{th2}.
\end{itemize}


\subsection{The cloning factor\label{sec:cfactor}}

In the physics literature an alternative estimator to $\Lambda_t^N$ \eqref{lambdatn} is often used, based on a concept called the `cloning factor' (see e.g.\ \cite{giardina,giardina2011simulating,hurtado19}). This is essentially a continuous-time jump process $(C^N_k (t) :t\geq 0)$ on $\R^+$ with $C^N_k (0)=1$, where at each selection event of size $n\geq -1$ at a given time $t$ the value is updated as
\begin{equation}\label{cldef}
C^N_k (t)=C^N_k (t-) (1+n/N)\ .
\end{equation}
Here $n=-1$ indicates a killing event, and $n\geq 0$ a cloning event according to the two parts of the generator \eqref{clonegen}. This idea comes from a branching process interpretation of the cloning ensemble related to the unnormalized measure $\nu_t^k$, since with 
\eqref{lkp} we have that
\[
\nu_t^k (1)\approx e^{\lambda_k t}\quad\mbox{as }t\to\infty\ ,
\]
so $\lambda_k$ corresponds to the volume expansion factor of the clone ensemble due to selection dynamics.

In our setting, the dynamics of $C^N_k (t)$ can be defined jointly with the cloning process via an extension of the generator \eqref{clonegen}
\[
\bar\LL_k^{N,clone} F(\underline{x},\zeta )=\bar\LL_k^{N,mc} F(\underline{x},\zeta)+\bar\LL_k^{N,-} F(\underline{x},\zeta )\ ,
\]
acting on functions that depend on the state $\underline{x}\in E^N$ and the cloning factor $\zeta\in\R^+$. With \eqref{mcgen} we have for cloning events
\begin{align*}
\bar\LL_k^{N,mc} F(\underline{x},\zeta ):=\sum_{i=1}^N \sum_{n=0}^N\frac{1}{{N\choose n}}&\sum_{A\subseteq\{ 1,..,N\}\atop |A|=n} \int_E W_k (x_i ,dy) p_{k,x_i }^N (n) \\
&\Big( F\big(\underline{x}^{A,x_i ;i,y},\zeta (1+n/N)\big) -F\big(\underline{x},\zeta\big)\Big)
\end{align*}
and with the third line of \eqref{filter1} for killing events
\[
\bar\LL_k^{N,-} F(\underline{x},\zeta ) =\sum_{i=1}^N\big(\VV_k(x_i )-c\big)^-\frac{1}{N}\sum_{j=1}^N \big( F(\underline{x}^{i,x_j},\zeta (1-1/N))-F(\underline{x} ,\zeta )\big)\ .
\]
So the joint process $\big( (\underline{X}_k^N (t),C_k^N (t)) :t\geq 0\big)$ is Markov, and observing only the cloning factor with the simple test function $G(\underline{x},\zeta )= \zeta$ we get
\begin{align}\label{cl1}
\bar\LL_k^{N,mc} G(\underline{x},\zeta )=&\sum_{i=1}^N \sum_{n=0}^N\frac{1}{{N\choose n}}\sum_{A\subseteq\{ 1,..,N\}\atop |A|=n} \int_E W_k (x_i ,dy) p_{k,x_i }^N (n) \zeta\cdot\frac{n}{N}\nonumber\\
=&\frac{\zeta}{N}\sum_{i=1}^N \big(\VV_k (x_i )-c\big)^+  .
\end{align}
In the last line we have used \eqref{pmean}, and in a similar fashion we get for killing events
\begin{equation}\label{cl2}
\bar\LL_k^{N,-} G(\underline{x},\zeta ) =-\frac{\zeta}{N}\sum_{i=1}^N\big(\VV_k(x_i )-c\big)^-  .
\end{equation}
Therefore
\[
\bar\LL_k^{N,clone} G(\underline{x},\zeta )=\frac{\zeta}{N}\sum_{i=1}^N \big(\VV_k(x_i ){-}c\big) = \zeta m^N (\underline{x}) (\VV_k)-\zeta c \ ,
\]
and analogously to \eqref{vte}, 
the expected time evolution of $C^N_k (t)$ is then given by
\begin{equation*}
    \frac{d}{dt}\E[C^N_k(t)]=\E[C^N_k(t)\cdot \mu^N_t(\VV_k-c)].
\end{equation*}
This is also the evolution of $\nu_t^N(e^{-tc})=e^{-tc}\nu_t^N(1)$, since
\begin{align*}
    \frac{d}{dt}\E[\nu_t^N(e^{-tc})]&=\E[\mu_t^N(\VV_k) e^{-tc}\nu_t^N(1)-c\ e^{-tc}\nu_t^N(1)]\\
    &=\E[\nu_t^N(e^{-tc})\cdot \mu^N_t(\VV_k-c)].
\end{align*}
With initial conditions $C_k^N (0)=1=\nu_t^N (1)$, a Gronwall argument analogous to \eqref{gronwall1} gives
\begin{equation*}
    \E[e^{tc}C^N_k(t)]=\E[\nu_t^N(1)]=\nu_t(1)\quad\mbox{for all }t\geq 0\mbox{ and }N\geq 1\ .
\end{equation*}
So $e^{tc}C^N_k(t)$ is an unbiased estimator for $\nu_t(1)$, which leads also to an alternative estimator for $\lambda_k(t)$ \eqref{lkp} given by
\begin{equation}\label{cflambda}
    \overline{\Lambda}_k^N(t):=\frac{1}{t}\log C^N_k(t)+c.
\end{equation}
Note that this is not itself unbiased as a consequence of the nonlinear transformation involving the logarithm.

Since $C_k^N (t)$ is defined as a product \eqref{cldef}, we can use another simple test function $G(\underline{x},\zeta )=\log\zeta$ to analyze the convergence behaviour of $\overline{\Lambda}_k^N(t)$. Analogously to \eqref{cl1} and \eqref{cl2} we get
\[
\bar\LL_k^{N,clone} G(\underline{x},\zeta )=\frac{1}{N}\sum_{i=1}^N \big(\VV_k(x_i ){-}c\big) = m^N (\underline{x}) (\VV_k)-c +O\Big(\frac{1}{N}\Big)\ ,
\]
where we have also used \eqref{pmean} and assumed that the  support of $p^N_{k,x_i}$ is bounded independently of $N$ (which is the case for common choices in the literature such as \eqref{pchoice}). This allows us to approximate $\log (1+n/N)=n/N +O(1/N^2 )$ as $N\to\infty$, leading to error terms of order $1/N$. 
Then, analogously to \eqref{marti} we get with $\log C_k^N (0)=0$ that
\begin{align*}
\MM_C^N (t):=&\log C_k^N (t) -\int_0^t \bar\LL_k^{N,clone} G(\underline{X}_k (s),C^N_k(s))\, ds\\
=&\log C_k^N (t) -t(\Lambda_k^N (t)-c)+t\, O\Big(\frac{1}{N}\Big)
\end{align*}
is a martingale. For the carr\'e du champ we obtain from a straightforward computation that
\[
\Gamma_k^{N,clone} S(\underline{x},\zeta )=\frac{1}{N^2}\sum_{i=1}^N\big|\VV_k(x_i )-c\big|
+O\Big(\frac{1}{N^2}\Big)\ ,
\]
and since the potential $\VV_k$ is bounded \eqref{vbound}, the quadratic variation of the martingale is bounded by
\[
\langle \MM_C^N \rangle (t)\leq \frac{t}{N}\big(\bar v_k +|c|\big)+O\Big(\frac{1}{N^2}\Big)\ .
\]
Therefore the estimator \eqref{cflambda} based on the cloning factor
\[
\bar\Lambda_k^N (t)=\frac{1}{t}\log C_k^N (t) +c=\Lambda_k^N (t)+O\Big(\frac{1}{N}\Big)+\frac{1}{t}\MM_C^N (t)
\]
is asymptotically equal to the basic estimator $\Lambda_k^N (t)$ \eqref{lambdatn}, with corrections that vanish as $1/N$ in the $L^p$-norm as $N\to\infty$ uniformly in $t\geq 0$, analogously to the discussion in Section \ref{sec:essential}. Therefore, the same convergence results as stated in the Theorem apply for $\bar\Lambda_k^N (t)$. Similar convergence results can be shown to hold for $e^{tc}C_k^N (t)$ as an estimator of $\nu_t (1)$ for fixed $t>0$, but naturally cannot hold uniformly in time. Since the object of interest is usually the long-time limit $\lambda_k$ \eqref{lkp}, the practical relevance of this is limited, in addition to the general point that random errors dominate the convergence as mentioned in Section \ref{sec:essential}. 
%
In practice, the basic ergodic average $\Lambda_k^N (t)$ \eqref{lambdatn} is more useful than the cloning factor in the application areas we have in mind. In particular, for alternative particle approximations such as \eqref{filter2} or \eqref{filter3} where cloning and killing events are effectively combined, it is not clear how to define a cloning factor, whereas $\Lambda_k^N (t)$ is always easily accessible.\\

\section{Efficiency and application of particle approximations\label{sec:ip}}

\subsection{Efficiency of algorithms}

Selection events (cloning or killing) in a particle approximation increase the correlations among the particles in the ensemble, 
and thereby decrease 
the resolution in the empirical distribution $\mu_t^N =\mu^N (\underline X_k (t))$, and ultimately the quality of the sample average in the estimator \eqref{lambdatn}. Therefore it is desirable to minimize the total rate $S_k (\underline{x})$ of selection events for a particle approximation. For algorithm \eqref{filter1} this is given by
\begin{equation}\label{sk1}
S_k^1 (\underline{x})=\sum_{i=1}^N \big| \VV_k (x_i )-c\big|\ ,
\end{equation}
and the same holds for the cloning algorithm \eqref{clonegen}, since the change in cloning rate is compensated exactly by the average number of clones created to obtain the same overall rate. It is easy to see that for a given state $\underline{x}$ of the clone ensemble, there is an optimal choice of $c$ to minimize this expression, given by the median of the fitness distributions $\VV_k (\underline x):=\big\{\VV_k (x_i ):i=1,\ldots ,N\big\}$. If the distribution of $\underline X_k (t)$ is unimodal with light enough tails, the median can be well approximated by the mean $\mu_t^N (\VV_k)$. Since both quantities can be computed with similar computational effort (or well approximated at reduced cost using only a subset of the ensemble), choosing
\[
c=c(t)=\mbox{median}\big(\VV_k (\underline X_k (t))\big)
\]
should be computationally optimal. In particular, the simplest choice $c=0$ in the cloning algorithm is in general far from optimal, so is choosing $c=\inf_{x\in E} \VV_k (x)$ to get rid of the killing part of the dynamics (see first remark in Section \ref{sec:cloning}).

Intuitively, algorithms \eqref{filter2} and \eqref{filter3} should lead to even lower total selection rates since every selection event increases the fitness potential, while in algorithms based on \eqref{filter1} it increases only on average and may also decrease as the result of selection events. 
Indeed for \eqref{filter2} we have
\begin{align}\label{sk2}
S_k^2 (\underline{x})&=\frac{1}{N}\sum_{i,j=1}^N \big( \VV_k (x_i )-\VV_k (x_j )\big)^+ =\frac{1}{2N} \sum_{i,j=1}^N \big| \VV_k (x_i )-\VV_k (x_j )\big| \nonumber\\
&\leq \frac{1}{2} \sum_{i=1}^N \Big(\big| \VV_k (x_i )-c\big| +\big|c-\VV_k (x_i )\big|\Big)= S_k^1 (\underline{x})\ ,
\end{align}
by symmetry of summations and the triangle inequality. The inequality is strict except for degenerate cases, e.g.\ if $\VV_k (x_i )$ takes only two values, and $c$ lies in between the two. In practice, in the scenarios which we have investigated, it turns out that unless the distribution of $\underline X_k (t)$ is seriously skewed, $S_k^2$ is strictly smaller than $S_k^1$ by a sizeable amount, as is illustrated later in Figure \ref{skfig} for the inclusion process. Algorithm \eqref{filter3} provides further improvement with
\begin{align}\label{sk3}
S_k^3 (\underline{x})&=\frac{1}{N}\sum_{i,j=1}^N \big( \VV_k (x_i )-\mu^N (\underline{x})(\VV_k )\big)^- \frac{\big( \VV_k (x_j )-\mu^N (\underline{x})(\VV_k )\big)^+}{\mu^N (\underline{x})\big( (\VV_k -\mu^N (\underline{x})(\VV_k ))^+ \big)}\nonumber\\
&=\frac12\sum_{i=1}^N \big| \VV_k (x_i )-\mu^N (\underline{x})(\VV_k )\big| \leq S_k^2 (\underline{x})\ .
\end{align}
Here we have used $\sum_{i=1}^N \big( \VV_k (x_i )-\mu^N (\underline{x})(\VV_k )\big) =0$ and Jensen's inequality to compare with $S_k^2 (\underline{x})$, since $v\mapsto |a-v|$ is convex for all $a\in\R$. 
Note that the rate of change of the mean fitness $\mu_t^N (\VV_k )$ is given by the same expression in all the above particle approximations,
\begin{equation}\label{muse}
\mu_t^N (\widehat{\LL}_k \VV_k) +\mu_t^N (\VV_k^2) -\mu_t^N (\VV_k)^2 \ .
\end{equation}
The first term due to mutation dynamics $\widehat{\LL}_k$ can have either sign and is identical in all algorithms, while the second due to selection is positive and given by the empirical variance of $\VV_k$. 
This follows from direct computations using the averaged test function $F(\underline x)=\mu^N (\underline x) (\VV_k )$ in \eqref{filter1}, \eqref{filter2}, \eqref{filter3} and \eqref{clonegen}, and is consistent with the evolution equation \eqref{muvolution}. So the mean fitness evolves until a mutation selection balance is reached and the rate of change \eqref{muse} vanishes, characterizing the stationary state of the particle approximation process. Note that this basic mechanism is identical in all particle approximations discussed here, so we expect the mean fitness to show a very similar behaviour. While finite size effects can lead to deviations also in the mean, the main difference between the algorithms is found on the level of variances and time correlations, which can be significantly reduced using \eqref{filter2} or \eqref{filter3} as illustrated in the next subsections. Since our main observable of interest $\Lambda_k^N (t)$ is an ergodic time average of $\mu_t^N (\VV_k)$, this can lead to significant improvements in the accuracy of the estimator \eqref{lambdatn}.
 
The correlations introduced by selection are counteracted by mutation dynamics, which occur independently for each particle and decorrelate the ensemble. 
The dynamics of correlation structures in cloning algorithms has been discussed in some detail recently in \cite{garrahan,nemoto2016,hidalgo18,hurtado19}, and can be understood in terms of ancestry in the generic population dynamics interpretation. Those results also discuss important non-ergodicity effects in the measurement of path properties and the interpretation of particle trajectories, which were already pointed out in \cite{giardina} and are also a subject of recent research \cite{limmer}. 
This poses interesting questions for rigorous mathematical investigations which are left to future work. Here we simply conclude with a numerical test in the next subsections, which supports the intuition that approximation \eqref{filter2} with minimal selection rates leads to variance reduction in the relevant estimators compared to the cloning algorithm. Since the selection rate in \eqref{filter2} depends on potential differences between pairs, implementation is in general more involved than for algorithms based on \eqref{filter1}. While the scaling $t N\log N$ of computational complexity with the size $N$ of the clone ensemble is the same, the prefactor and computational cost in practice may be higher and this has to be traded off against gains in accuracy on a case by case basis. For the examples studied below we find a computationally efficient implementation of \eqref{filter2} providing a clear improvement over the standard cloning algorithm, which is the main contribution of this paper in this context. Algorithm \eqref{filter3}, on the other hand, provides only marginal improvement over \eqref{filter2}, but cannot be implemented as efficiently in our area of interest.

\subsection{Current large deviations for lattice gases}

In the following we consider one-dimensional stochastic lattice gases with periodic boundary conditions on the discrete torus $\T_L$ with $L$ sites and a fixed number of particles $M$. Within our general framework, they are simply Markov chains on the finite state space $E$ of all particle configurations, which have been of recent research interest in the context of current fluctuations. We denote configurations by $\eta =(\eta_x :x\in\T_L )$ where $\eta_x \in\N_0$ is interpreted as the mass (or number of monomers) at site $x$, and the process is denoted as $(\eta (t):t\geq 0)$. In order to use standard notation for lattice gases, in this and the following subsection we change notation, and in particular the use of $x,y\in\T_L$ is different to the use of those symbols in previous sections where they denoted states in $E$.  Monomers jump to nearest neighbour sites with rates $u(\eta_x ,\eta_y )\geq 0$ for $y=x\pm 1$ depending on the occupation numbers of departure and target site, multiplied with a spatial bias $p=1-q\in [0,1]$. 
The generator is of the form
\begin{align}\label{ipgen}
\LL f(\eta )=\sum_{x\in\T_L} \Big[ &p\, u(\eta_x ,\eta_{x+1}) \big( f(\sigma_{x,x+1}\eta)-f(\eta )\big)\nonumber\\
& +q\, u(\eta_x ,\eta_{x-1}) \big( f(\sigma_{x,x-1}\eta)-f(\eta )\big)\Big]\ ,
\end{align}
where $\sigma_{x,y}\eta$ results from the configuration $\eta$ after moving one particle from $x$ to $y$. The number of particles $M=\sum_{x\in\T_L} \eta_x$ is a conserved quantity, but otherwise we assume the process to be irreducible for any fixed $M$, which is ensured for example by positivity of the rates, i.e. for all $k,l\geq 0$
\[
u(k,l)=0\quad\Leftrightarrow\quad k=0\ .
\]
This class includes various models that have been studied in the literature, for example the inclusion process introduced in \cite{giardina09}, where
\begin{equation}\label{iprates}
u(k ,l )=k (d+l)\quad\mbox{for all }k,l\geq 0\ ,
\end{equation}
with a positive parameter $d>0$. Particles perform independent jumps with rate $d$ and in addition are attracted by each particle on the target site with rate $1$, giving rise to the `inclusion' interaction. 
This model has attracted recent attention due the presence of condensation phenomena \cite{grosskinsky13,bianchi17} and in the context of large deviations of the particle current \cite{pizzo2}, and we will use this as an example in Section \ref{sec:iip}. 
Other well-studied models covered by our set-up are the exclusion process with state space $E\subset\{ 0,1\}^{\T_L}$ and $u(\eta_x ,\eta_y )=\eta_x (1-\eta_y )$, or zero-range processes with $E\subset\N_0^{\T_L}$ and rates $u(\eta_x ,\eta_y )=u(\eta_x )$ depending only on the occupation number on the departure site. 

In terms of previous notation, the jump rates for a lattice gas of type \eqref{ipgen} between any two configurations $\eta$ and $\zeta$ are given as
\begin{equation}\label{mura}
W(\eta ,\zeta )=\sum_{x\in\T_L} \Big( p\, u(\eta_x ,\eta_{x+1} )\delta_{\zeta ,\sigma_{x,x+1} \eta }+q\, u(\eta_x ,\eta_{x-1} )\delta_{\zeta ,\sigma_{x,x-1} \eta }\Big)\ .
\end{equation}
In the following we focus on lattice gases where $\sum_{x} u(\eta_x ,\eta_{x+1}) =\sum_{x} u(\eta_x ,\eta_{x-1})$ for all configurations $\eta$. While this is not true in general for models of type \eqref{ipgen}, it holds for many examples including inclusion, exclusion and zero-range processes mentioned above. 
With $p+q=1$, the total exit rate out of configuration $\eta$ is then simply given by
\begin{equation}\label{wcond}
w(\eta )=\sum_{x\in\T_L} \Big( p\,u(\eta_x ,\eta_{x+1}) +q\,u(\eta_x ,\eta_{x-1})\Big) =\sum_{x\in\T_L} u(\eta_x ,\eta_{x+1})\ .
\end{equation}
We are interested in an observable $A_T$ measuring the total particle current up to time $T$, which is achieved by choosing  $h(\eta )\equiv 0$ in \eqref{oss_jump} and
\[
g(\eta ,\zeta )=\pm 1\quad\mbox{if }\zeta =\sigma_{x,x\pm 1} \eta\quad\mbox{and}\quad g(\eta ,\zeta )=0\quad\mbox{otherwise}\ .
\]
Using \eqref{wcond} we see by direct computation that the potential \eqref{vk} takes the simple form
\begin{equation}\label{vvkip}
\VV_k (\eta )=(Q_k -1)w(\eta )\quad\mbox{where}\quad Q_k 
:=pe^k +qe^{-k} \ .
\end{equation}
Modified mutation rates $W_k (\eta ,\zeta )$ are given by \eqref{mura} replacing $(p,q)$ by $(pe^k ,qe^{-k} )$, leading to modified total exit rates
\begin{equation}\label{wkip}
w_k (\eta ) =Q_k \sum_{x\in\T_L} u(\eta_x ,\eta_{x+1})=Q_k w(\eta )\ .
\end{equation}
The similarity of $\VV_k$ and $w_k$ for lattice gases \eqref{ipgen} that obey \eqref{wcond} provides a direct relation between mutation and selection rates, and allows us to set up an efficient rejection based implementation of a particle approximation $(\underline\eta_k (t):t\geq 0)$ based on the efficient algorithm \eqref{filter2}. In the following we omit the subscript $k$ for configurations and write $\underline\eta (t)=(\eta^i (t),i=1,\ldots ,N)$ to simplify notation. For given parameters $p,q=1-p$ and fixed $k\in\R$ we distinguish two cases.\\

\noindent \textbf{$\mathbf{Q_k <1}$.} We sample the ensemble of $N$ clones at a total rate of $\WW (\underline\eta ):=\sum_{i=1}^N w(\eta^i )$, and pick a clone $i$ with probability $w (\eta^i )/\WW (\underline\eta )$ for the next event. With probability $Q_k \in (0,1)$ this is a simple mutation within clone $i$, and then we replace $\eta^i$ by $\zeta^i$ with probability $W_k (\eta^i ,\zeta^i )/w_k (\eta^i )$. Otherwise, with probability $1-Q_k$ we perform a selection event following the second line in \eqref{filter2}: Pick a clone $j$ uniformly at random (including $i$). If
\[
\VV_k (\eta^j )>\VV_k (\eta^i )\quad\mbox{or equivalently}\quad w(\eta^j )<w(\eta^i )
\]
(with \eqref{vvkip} and since $Q_k <1$), replace $\eta^i$ by $\eta^j$ with probability $\big( w(\eta^i )-w(\eta^j )\big) /w(\eta^i )$. This procedure ensures that mutation and selection events are sampled with the correct rates as required in \eqref{filter2}.\\

\noindent\textbf{$\mathbf{Q_k >1}$.} We sample the ensemble of $N$ clones at a total rate of $Q_k \WW (\underline\eta )$, and pick a clone $i$ with probability $w (\eta^i )/\WW (\underline\eta )$ and a clone $j$ uniformly at random. If $w(\eta^j )<w(\eta^i )$ we replace $\eta^j$ by $\eta^i$ with probability $\frac{Q_k -1}{Q_k}\frac{w(\eta^i )-w(\eta^j )}{w(\eta^i )}$. Then we mutate clone $i$ as above, combining the mutation and selection event as in the cloning algorithm.\\

\noindent\textbf{Remarks.} Note that $Q_k =1$ is equivalent to $k=0$, which corresponds to the original process with $\lambda_0 =0$ and does not require any estimation. 
For $Q_k >1$ we perform mutation and selection events simultaneously, in analogy to the cloning procedure explained in Section \ref{sec:cloning}, but can use the efficient algorithm \eqref{filter2}. 
For $Q_k <1$ no mutation or selection event occurs with probability $(1-Q_k )\frac{w(\eta^j )}{w(\eta^i )} \1 (w(\eta^j )<w(\eta^i ))$, and a high rate of such rejections is not desirable for computational efficiency. But even for very small values of $Q_k$ the second factor is usually significantly smaller than $1$ (or simply $0$), since clone $i$ was picked with probability proportional to $w(\eta^i )$ and $j$ uniformly at random.

Note also that if the cloning algorithm \eqref{clonegen} is implemented with the common choice $c=0$ for a lattice gas of the type discussed here, due to \eqref{vvkip} and \eqref{wkip} the average number of clones per event \eqref{pmean} is
\[
m_k^N (\eta^i )=\big(\VV_k (\eta^i )\big)^+ /w_k (\eta^i )=\frac{Q_k -1}{Q_k}\in (0,1)\quad\mbox{if }Q_k >1\ ,
\]
and $0$ for $Q_k <1$, where only killing occurs. In particular, this is independent of the state $\underline\eta$ of the clone ensemble, and the standard distribution of the form \eqref{pchoice} is a simple Bernoulli random variable.

\begin{figure}
\begin{center}
\includegraphics[width=0.45\textwidth]{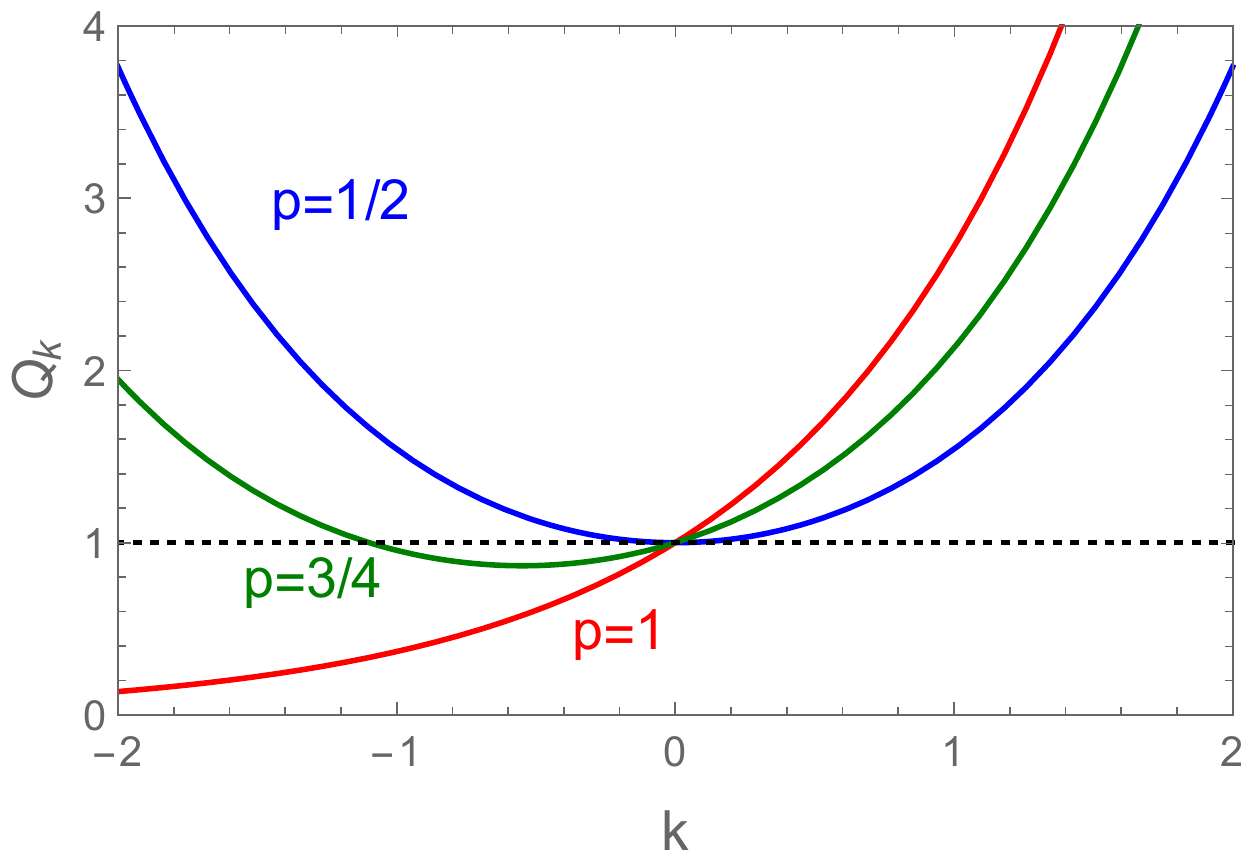}\quad\includegraphics[width=0.45\textwidth]{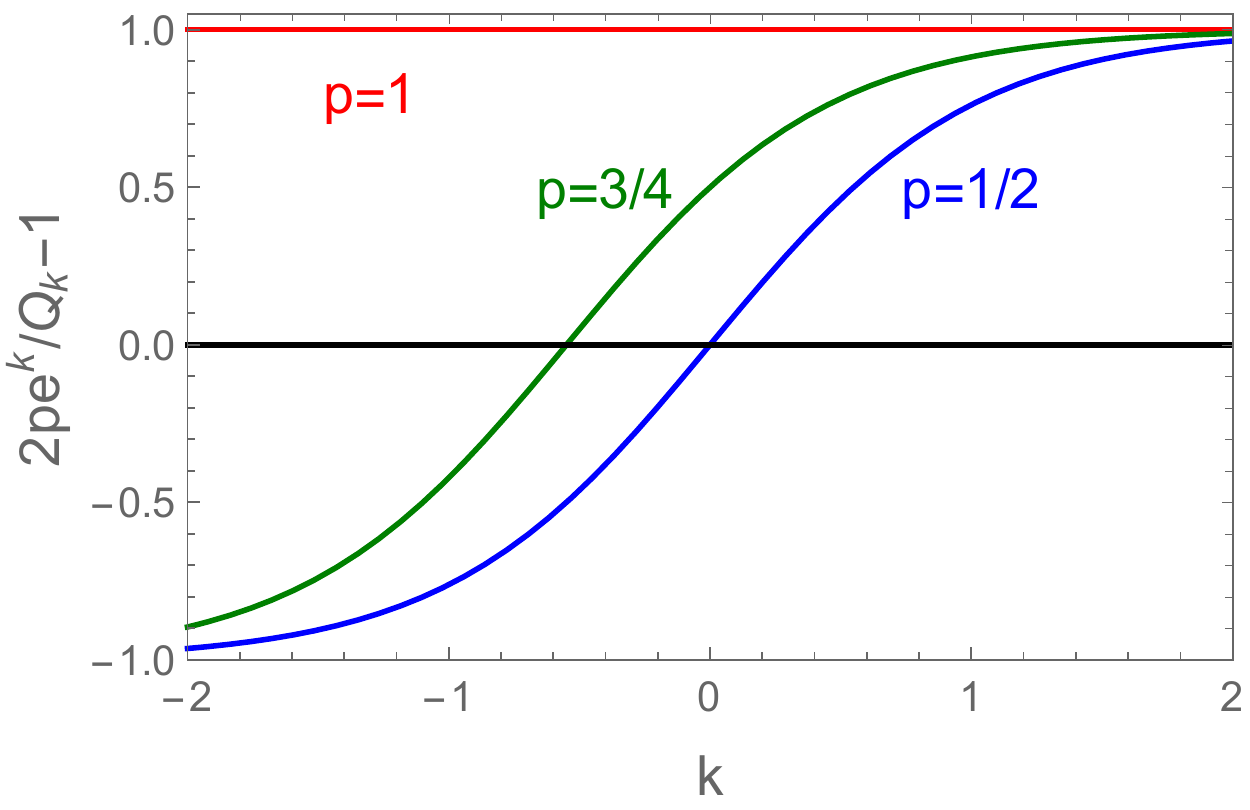}
\end{center}
\caption{\label{fig:qk} Illustration of $Q_k$ (left) as given in \eqref{vvkip} and the drift $2pe^k /Q_k-1$ for the modified dynamics (right) as a function of $k$ for different values of the asymmetry $p=1-q$. The minimum of $Q_k$ is $2\sqrt{pq}$, attained at $k=\frac12\log\frac{q}{p}\in [-\infty ,\infty ]$, which is also where the modified drift vanishes.
}
\end{figure}

While with \eqref{wkip} the total mutation rate is $Q_k \WW (\underline\eta )$, selection rates \eqref{sk1}, \eqref{sk2} and \eqref{sk3} can be written as
\begin{align}\label{sks}
S_k^1 (\underline\eta )&=\sum_{i=1}^N \big| (Q_k -1)w(\eta^i )-c\big|\stackrel{c=0}{=} |Q_k-1|\WW (\underline\eta )\nonumber\\
S_k^2 (\underline\eta )&=|Q_k-1|\frac{1}{2N} \sum_{i,j=1}^N \big| w(\eta^i )-w(\eta^j)\big|\nonumber\\
S_k^3 (\underline\eta )&=|Q_k-1| \frac12\sum_{i=1}^N \big| w(\eta^i )-\mu^N (\underline\eta )(w)\big|\ .
\end{align}
So for very small values of $Q_k$ close to $0$ the mutation rate can become very small in comparison to selection, which means that significant computation time is devoted to re-weighting by selection, rather than advancing the dynamics via mutation events. This effect is typically much stronger for the standard cloning algorithm with $c=0$, and occurs for example for totally asymmetric lattice gases with $p=1$ and negative $k$ conditioning on low currents. In Figure \ref{fig:qk} we include a sketch of $Q_k$ for different values of asymmetry, including also the drift of the modified dynamics, which can be reversed in partially asymmetric systems.

\begin{figure}
\begin{center}
\includegraphics[width=0.6\textwidth]{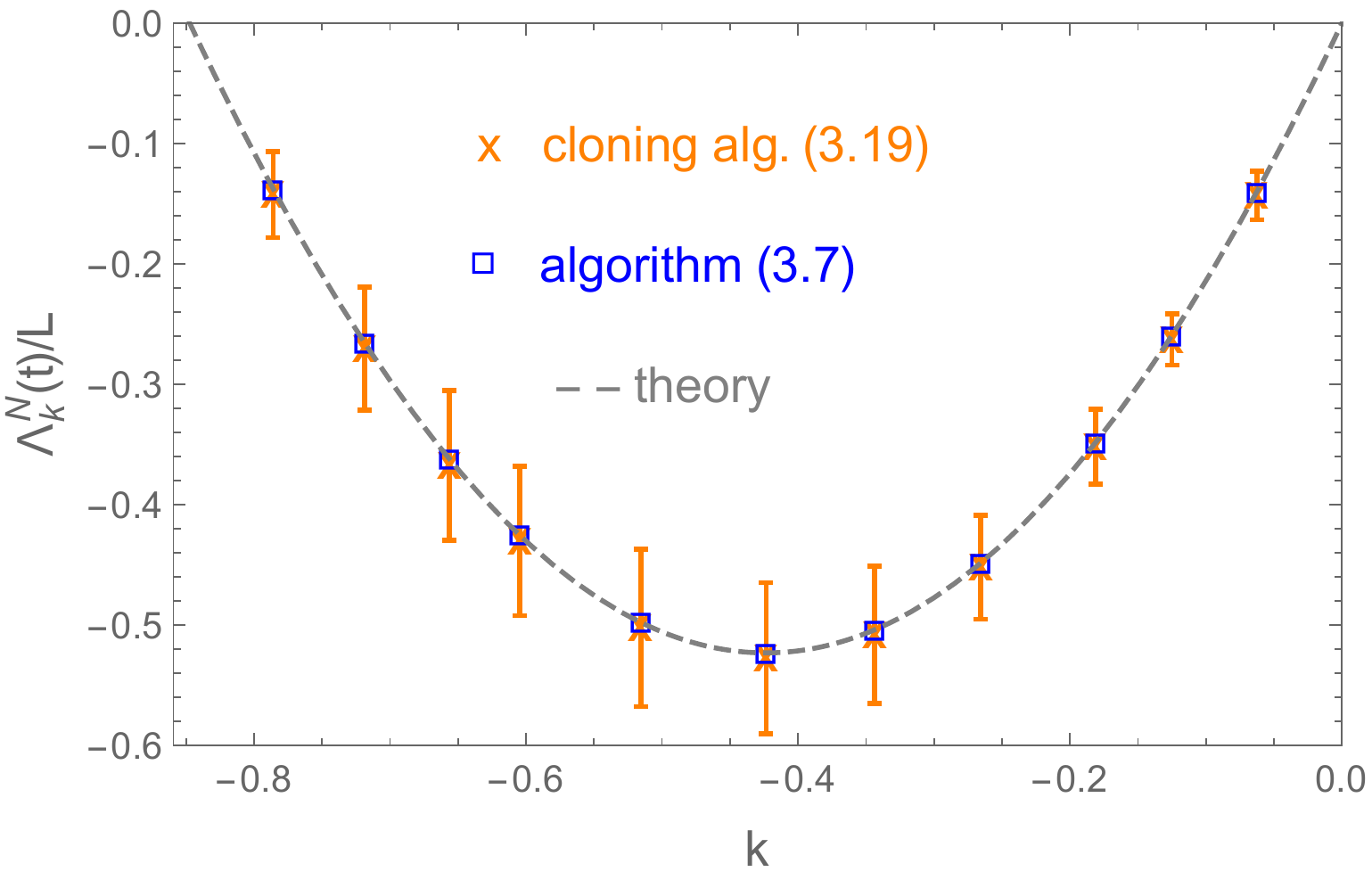}\\[5mm]
\includegraphics[width=0.45\textwidth]{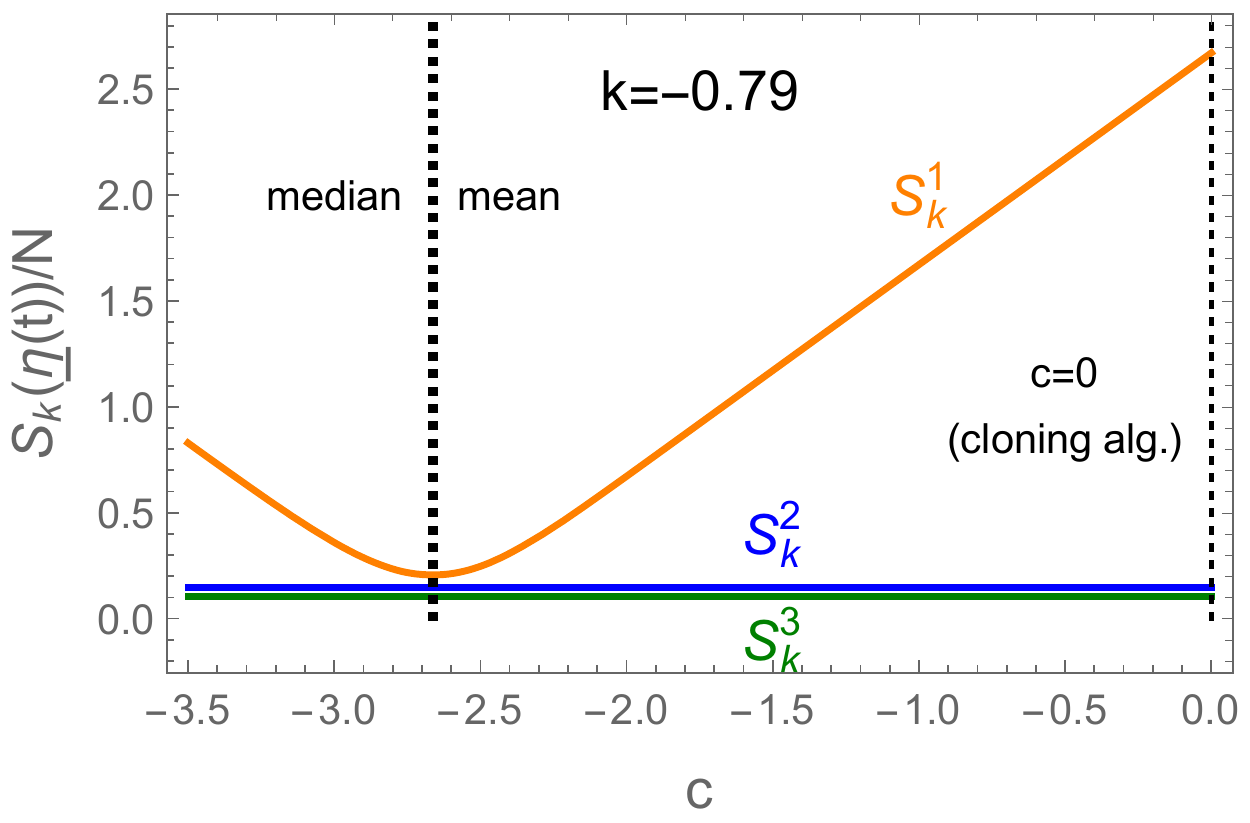}\quad\includegraphics[width=0.45\textwidth]{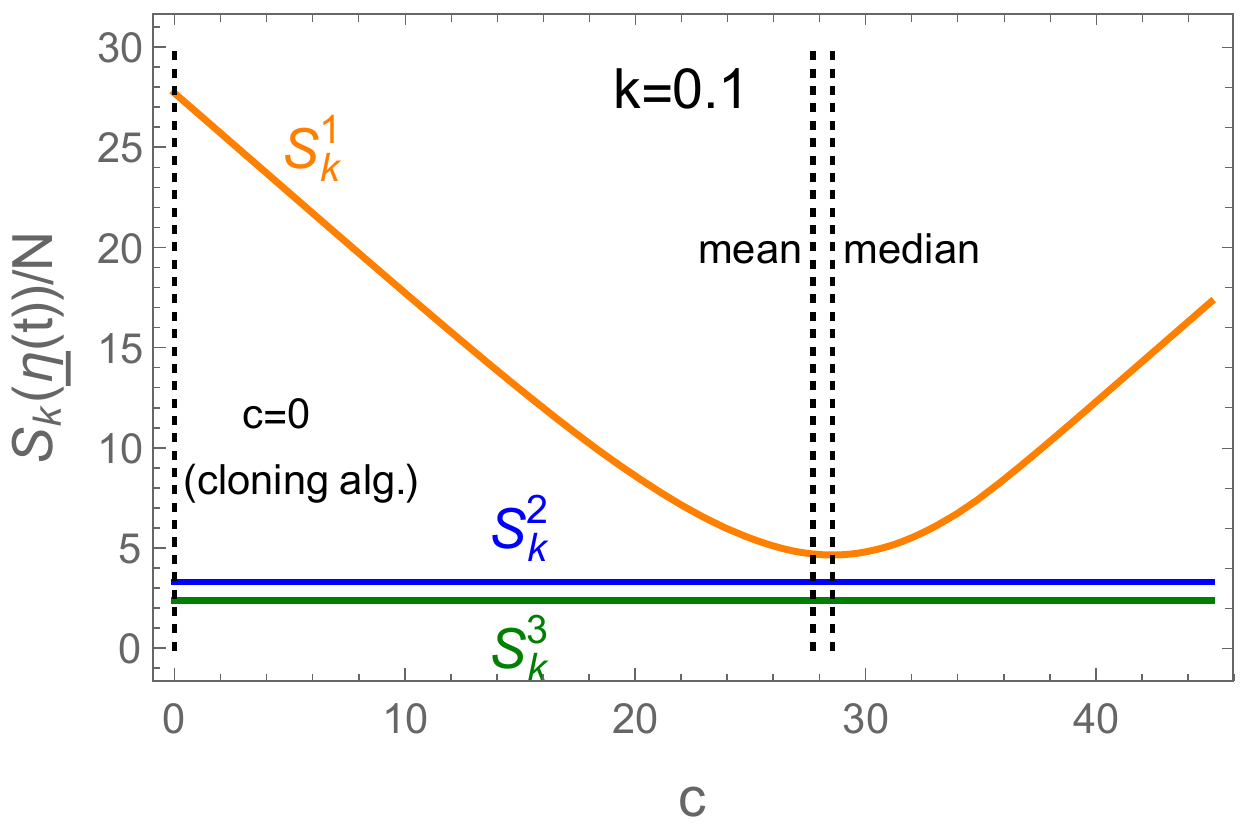}
\end{center}
\caption{\label{skfig}
Inclusion process \eqref{iprates} with $d=1$, system size $L=64$, $M=128$ particles, asymmetry $p=0.7$ and $N=2^{11}$ clones at time $t=42000$. 
(Top) The rescaled estimator $\Lambda_k^N (t)/L$ as a function of $k$ in the convergent regime, comparing the cloning algorithm \eqref{clonegen} with $c=0$ (orange) and algorithm \eqref{filter2} (blue). Error bars indicate $5$ standard deviations, which are bounded by the size of the symbols for \eqref{filter2}.
(Bottom) Illustration of the relationship between $S_k^1$ depending on $c$, and $S_k^2$ and $S_k^3$ \eqref{sks} for $k=-0.79$ (left) and $k=0.1$ (right) based on the state $\underline{\eta}(t)$ of the clone ensemble.
}
\end{figure}

\begin{figure}
\begin{center}
\includegraphics[width=0.45\textwidth]{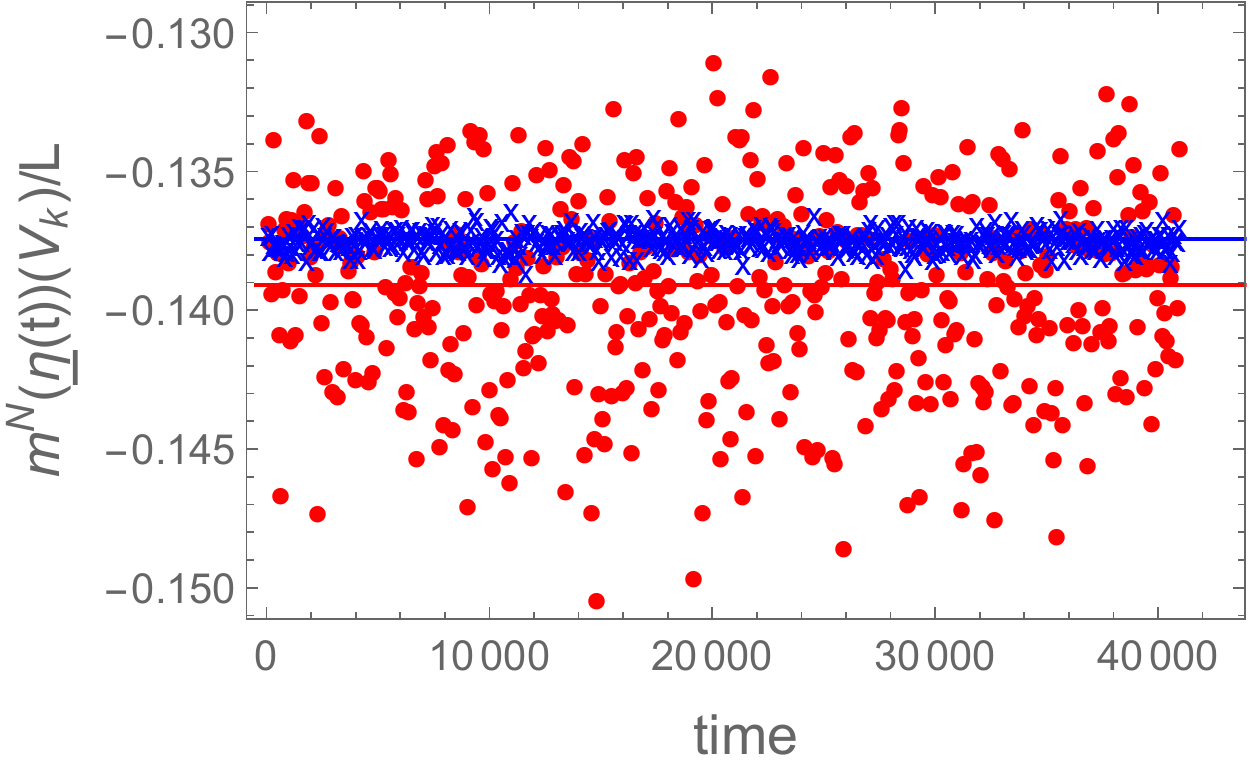}\quad\includegraphics[width=0.45\textwidth]{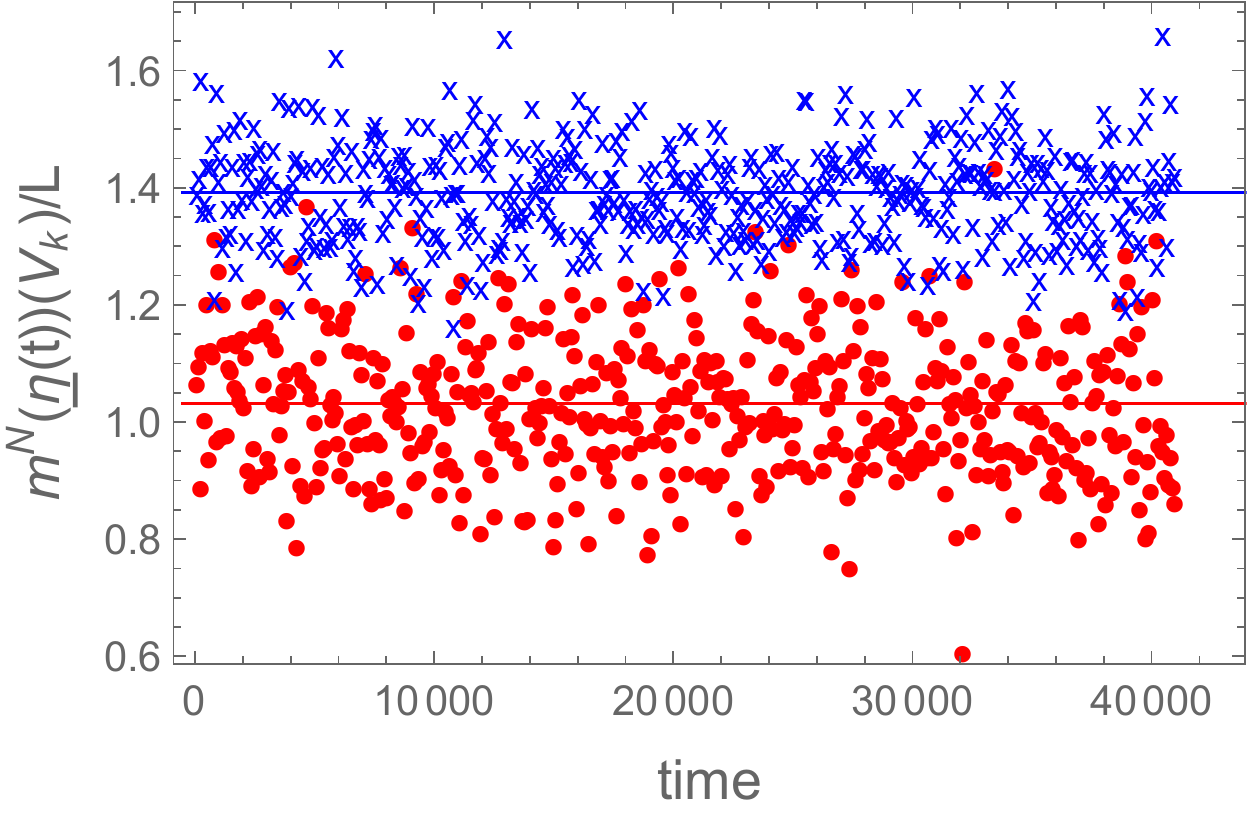}
\end{center}
\caption{\label{seriesfig}
Inclusion process \eqref{iprates} with $d=1$, system size $L=64$, $M=128$ particles, asymmetry $p=0.7$ and $N=2^{11}$ clones. Time series of the mean fitness $m^N (\underline\eta (t))(\VV_k)/L$ for the cloning algorithm (red dots) and algorithm \eqref{filter2} (blue crosses), with time averages indicated by full lines. (Left) In the convergent regime for $k=-0.79$ we see a clear variance reduction using \eqref{filter2} with similar time average. (Right) In the divergent regime for $k=0.1$ we have similar variance but \eqref{filter2} improves on the time average.
}
\end{figure}

In Figure \ref{skfig} we compare the cloning algorithm to algorithms \eqref{filter2} and \eqref{filter3}
for an inclusion process with $d=1$, $L=64$, $M=128$ and asymmetry $p=0.7$. It is known \cite{pizzo2} that the SCGF $\lambda_k$ scales linearly with the system size $L$, and outside the convergent regime $k\in [-\ln (\frac{1-p}{p}),0] \approx [-0.85,0]$ the rescaled SCGF $\lambda_k /L$ diverges as $L\to\infty$ (divergent regime). We compare estimates $\Lambda_k^N (t)$ for the cloning algorithm \eqref{clonegen} with $c=0$ and algorithm \eqref{filter2} in the convergent regime. We use initial conditions where $M$ particles are distributed on $L$ lattice sites uniformly at random, and a burn-in time of $10\cdot L=640$ as discussed in \eqref{lkap} and \eqref{lkat}. This leads to an obvious adaption of the integration interval in the estimator $\Lambda_k^N (t)$ \eqref{lambdatn}, but we do not alter the notation here to keep it simple. Both algorithms perform very well and agree with a simple theoretical estimate based on bias reversal, which is not the main concern in this paper and we refer the reader to \cite{pizzo2}. Enlarged error bars indicating $5$ standard deviations reveal that \eqref{filter2} is significantly more accurate than \eqref{clonegen}. 
This is due to lower total selection rates $S_k$ illustrated at the bottom in the converging and diverging regime. While $S_k^2$ for \eqref{filter2} is much lower than $S_k^1$ with $c=0$, $S_k^3$ does not offer significant further improvement. Since the efficient rejection based implementation of \eqref{filter2} explained above does not work for \eqref{filter3}, we focus on \eqref{filter2} in our context. 
The much higher selection rate for the cloning algorithm with $c=0$ leads to a significantly higher time variation of the average potential in the convergent regime compared to algorithm \eqref{filter2}, as is illustrated in Figure \ref{seriesfig}. So in comparison to standard cloning, algorithm \eqref{filter2} leads to reduced finite size effects and/or a significant variance reduction in this example, and a significant improvement of convergence of the estimator \eqref{lambdatn}. We have checked that this also holds for zero-range processes with bounded rates. 
These promising first numerical results pose interesting questions for a systematic study of practical properties of the algorithms and associated time correlations for future work, also in comparison with various recent results on improvements of cloning algorithms \cite{nemoto2016,brewer,ferre}.

\subsection{Details for the inclusion process\label{sec:iip}}

We summarize the procedure outlined in the previous subsection for the inclusion process with rates \eqref{iprates} on the torus $\T_L$ with $M$ particles in pseudo-code given below. Besides fixing the model parameter $d>0$ and the tilt $k\in\R$, the specific parameters for the estimator are the ensemble size $N$ and the total simulation time $t$, which lead to the estimator $\Lambda_k^N (t)$ as given in \eqref{lambdatn}. For simplicity we do not include any burn-in times in this description, which would obviously be used in practice. In this implementation we make a further simplification which is very common for continuous-time jump dynamics of large systems: We replace exponentially distributed random time increments by their expectation, given by $\Delta t=1/\WW (\underline\eta )$ for $Q_k <1$ and $\Delta t=1/(Q_k \WW (\underline\eta ))$ for $Q_k >1$.  
Since with \eqref{vvkip}
\[
\frac{1}{N}\sum_{i=1}^N \VV_k (\eta^i (s)) 
=\frac{Q_k -1}{N} \WW (\underline\eta )\ ,
\]
we get for increments in the evaluation of the ergodic time integral in \eqref{lambdatn}
\[
\Delta t\frac{1}{N}\sum_{i=1}^N \VV_k (\eta^i (s))=\left\{\begin{array}{cl} \frac{Q_k -1}{N}&,\ Q_k <1\\ \frac{Q_k -1}{Q_k N}&, Q_k >1\end{array}\right.\ .
\]
These are independent of the actual state $\underline\eta$ of the clones, so evaluation of $\Lambda_k^N (t)$ in \eqref{lambdatn} can be achieved by a simple integer counter $\hat\Lambda_k^N$ as explained in the pseudocode Algorithm \ref{alg1} and \ref{alg2}. While this counter may appear similar to the cloning factor explained in Section \ref{sec:cfactor} at first glance, we want to stress that here finer increments of $+1$ are added after every event (not only selections).

\begin{algorithm}
\caption{IP \eqref{ipgen} with rates \eqref{iprates} and $Q_k <1$ \eqref{vvkip}\label{alg1}}
\textbf{Parameters} $N$ number of clones; $t$ simulation time\;
\textbf{Initialize} configurations\ $\eta^i$, $i=1,\ldots ,N$ with mass $M$ each\;
$w(\eta^i )=d\, M+\sum_x \eta^i_x \eta^i_{x+1}$; $\WW (\underline\eta )=\sum_j w(\eta^j )$; $s=\frac{1}{\WW  (\underline\eta )}$; $\hat\Lambda_k^N =1$\;
\smallskip
 \While{$s<t$}{
  pick clone $i$ with probability $w(\eta^i )/\WW (\underline\eta )$\;
  \eIf{$R\sim U(0,1)<Q_k$}{
   $\eta^i \leftarrow \zeta$, $\zeta$ chosen with probability\ $\frac{W_k (\eta^i ,\zeta )}{Q_k w(\eta^i )}$\hfill (mutation)\;
   }{
   pick clone $j$ uniformly at random\;
   \If{$w(\eta^j )<w(\eta^i )$
   }{
   $\eta^i \leftarrow\eta^j$ with probability $\frac{w(\eta^i )-w(\eta^j )}{w(\eta^i )}$\hfill (selection)\;
   }
  }
 $s\leftarrow s+\frac{1}{\WW (\underline\eta )}$; $\hat\Lambda_k^N \leftarrow \hat\Lambda_k^N +1$\;
 }
 \textbf{Output} $\Lambda_k^N (t) =\frac{Q_k -1}{N t}\,\hat\Lambda_k^N$\;
\end{algorithm}

\begin{algorithm}
\caption{IP \eqref{ipgen} with rates \eqref{iprates} and $Q_k >1$ \eqref{vvkip}\label{alg2}}
\textbf{Parameters} $N$ number of clones; $t$ simulation time\;
\textbf{Initialize} configurations\ $\eta^i$, $i=1,\ldots ,N$ with mass $M$ each\;
$w(\eta^i )=d\, M+\sum_x \eta^i_x \eta^i_{x+1}$; $\WW (\underline\eta )=\sum_j w(\eta^j )$; $s=\frac{1}{Q_k \WW}$; $\hat\Lambda_k^N =1$\;
\smallskip
 \While{$s<t$}{
   pick clone $i$ with probability $w(\eta^i )/\WW (\underline\eta )$\;
   pick clone $j$ uniformly at random\;
   \If{$w(\eta^j )<w(\eta^i )$}{
   $\eta^j \leftarrow\eta^i$ with probability $\frac{Q_k -1}{Q_k}\frac{w(\eta^i )-w(\eta^j )}{w(\eta^i )}$\hfill (selection)\;
   }
   $\eta^i \leftarrow \zeta$, $\zeta$ chosen with probability $\frac{W_k (\eta^i ,\zeta )}{Q_k w(\eta^i )}$\hfill (mutation)\;
 $s\leftarrow s+\frac{1}{Q_k \WW (\underline\eta )}$; $\hat\Lambda_k^N \leftarrow\hat\Lambda_k^N +1$\;
 }
 \textbf{Output} $\Lambda_k^N (t) =\frac{Q_k -1}{Q_k N t}\,\hat\Lambda_k^N$\;
\end{algorithm}

\section{Discussion}

We have presented an analytical approach to cloning algorithms based on McKean interpretations of Feynman-Kac semigroups that have been introduced in the applied probability literature. This allows us to establish rigorous error bounds for the cloning algorithm in continuous time, and to suggest a more efficient variant of the algorithm which can be implemented effectively for current large deviations in stochastic lattice gases. The latter is based on minimizing the selection rate in a standard population dynamics interpretation of particle approximations of non-linear processes. We include a first application of this idea in the context of inclusion processes, but its full potential will be explored in future more systematic studies of optimization of cloning-type algorithms. 
The rigorous results fully reported in \cite{mathpaper} apply under very general conditions, demanding bounded jump rates and existence of a spectral gap for the underlying jump process. These impose no restriction for lattice gases with a fixed number of particles, which are essentially finite state Markov chains. We anticipate that these techniques can also be applied for more general processes including diffusive, piecewise deterministic, or possibly non-Markovian dynamics (see \cite{cavallaro} for first heuristic results in this direction). Another interesting direction would be a rigorous analysis of the detailed ergodic properties of trajectories in the clone ensemble based on recent results in \cite{nemoto2016,hidalgo18,limmer}.

\section*{Acknowledgements}
\noindent This work was supported by The Alan Turing Institute under the EPSRC grant EP/N510129/1 and The Alan Turing
Institute--Lloyds Register Foundation Programme on Data-centric
Engineering.\\
AP acknowledges support by the National Group of Mathematical Physics (GNFM-INdAM), and by Imperial College together with the Data Science Institute and Thomson-Reuters Grant No.\ 4500902397-3408.


\begin{thebibliography}{10}

\bibitem{anderson}
James~B. Anderson.
\newblock A random-walk simulation of the {S}chr{\"o}dinger equation:
  ${H}^+_3$.
\newblock {\em Journal of Chemical Physics}, 63:1499, 1975.

\bibitem{grassberger}
Peter Grassberger.
\newblock Go with the winners: a general {M}onte {C}arlo strategy.
\newblock {\em Computer Physics Communications}, 147(1):64--70, 2002.

\bibitem{giardina}
Cristian Giardina, Jorge Kurchan, and Luca Peliti.
\newblock Direct evaluation of large-deviation functions.
\newblock {\em Physical review letters}, 96(12):120603, 2006.

\bibitem{lecomte2007numerical}
Vivien Lecomte and Julien Tailleur.
\newblock A numerical approach to large deviations in continuous time.
\newblock {\em Journal of Statistical Mechanics: Theory and Experiment},
  2007(03):P03004, 2007.

\bibitem{giardina2011simulating}
Cristian Giardina, Jorge Kurchan, Vivien Lecomte, and Julien Tailleur.
\newblock Simulating rare events in dynamical processes.
\newblock {\em Journal of Statistical Physics}, 145(4):787--811, 2011.

\bibitem{sollich}
Robert~L. Jack and Peter Sollich.
\newblock Large deviations and ensembles of trajectories in stochastic models.
\newblock {\em Progress of Theoretical Physics Supplement}, 184:304--317, 2010.

\bibitem{nemoto2016}
Takahiro Nemoto, Freddy Bouchet, Robert~L. Jack, and Vivien Lecomte.
\newblock Population-dynamics method with a multicanonical feedback control.
\newblock {\em Physical Review E}, 93:062123, 2016.

\bibitem{hidalgo18}
Esteban~Guevara Hidalgo.
\newblock {\em Cloning Algorithms: from Large Deviations to Population
  Dynamics}.
\newblock Ph.d. thesis, Universit{\'e} Sorbonne Paris Cit{\'e} ---
  Universit{\'e} Paris Diderot 7, 2018.

\bibitem{hidalgo1}
Takahiro Nemoto, Esteban Guevara~Hidalgo, and Vivien Lecomte.
\newblock Finite-time and finite-size scalings in the evaluation of
  large-deviation functions: Analytical study using a birth-death process.
\newblock {\em Physical Review E}, 95:012102, 2017.

\bibitem{hidalgo2}
Esteban Guevara~Hidalgo, Takahiro Nemoto, and Vivien Lecomte.
\newblock Finite-time and finite-size scalings in the evaluation of
  large-deviation functions: Numerical approach in continuous time.
\newblock {\em Physical Review E}, 95:062134, 2017.

\bibitem{ferre}
Gr{\'e}goire Ferr{\'e} and Hugo Touchette.
\newblock Adaptive sampling of large deviations.
\newblock {\em Journal of Statistical Physics}, 172(6):1525--1544, 2018.

\bibitem{brewer}
Tobias Brewer, Stephen~R Clark, Russell Bradford, and Robert~L Jack.
\newblock Efficient characterisation of large deviations using population
  dynamics.
\newblock {\em Journal of Statistical Mechanics: Theory and Experiment},
  2018(5):053204, 2018.

\bibitem{hurtado19}
Carlos P{\'e}rez-Espigares, Pablo I.~Hurtado.
\newblock Sampling rare events across dynamical phase transitions.
\newblock arXiv:1902.01276

\bibitem{del2000moran}
Pierre Del~Moral and Laurent Miclo.
\newblock A {M}oran particle system approximation of {F}eynman-{K}ac formulae.
\newblock {\em Stochastic Processes and their Applications}, 86(2):193--216,
  2000.

\bibitem{del2000branching}
Pierre Del~Moral and Laurent Miclo.
\newblock Branching and interacting particle systems approximations of
  {F}eynman-{K}ac formulae with applications to non-linear filtering.
\newblock In {\em Seminaire de probabilites XXXIV}, pages 1--145. Springer,
  2000.

\bibitem{del2004feynman}
Pierre Del~Moral.
\newblock {\em Feynman-Kac Formulae}.
\newblock Springer, 2004.

\bibitem{rousset2006control}
Mathias Rousset.
\newblock On the control of an interacting particle estimation of
  {S}chr{\"o}dinger ground states.
\newblock {\em SIAM Journal on Mathematical Analysis}, 38(3):824--844, 2006.

\bibitem{mathpaper}
Letizia Angeli, Stefan Grosskinsky, and Adam~M. Johansen.
\newblock Limit theorems for cloning algorithms.
\newblock under review (arxiv:1902.00509)

\bibitem{giardina09}
Cristian Giardin{\`a}, Jorge Kurchan, Frank Redig, and Kiamars Vafayi.
\newblock Duality and hidden symmetries in interacting particle systems.
\newblock {\em Journal of Statistical Physics}, 135(1):25--55, 2009.

\bibitem{pizzo2}
Paul Chleboun, Stefan Grosskinsky, and Andrea Pizzoferrato.
\newblock Current large deviations for partially asymmetric particle systems on
  a ring.
\newblock {\em Journal of Physics A: Mathematical and Theoretical},
  51(40):405001, 2018.

\bibitem{lazarescu}
Alexandre Lazarescu.
\newblock The physicist's companion to current fluctuations: one-dimensional
  bulk-driven lattice gases.
\newblock {\em Journal of Physics A: Mathematical and Theoretical},
  48(50):503001, 2015.

\bibitem{hurtado2014}
Pablo~I. Hurtado, Carlos~P. Espigares, Jes{\'u}s~J. del Pozo, and Pedro~L.
  Garrido.
\newblock Thermodynamics of currents in nonequilibrium diffusive systems:
  Theory and simulation.
\newblock {\em Journal of Statistical Physics}, 154(1):214--264, 2014.

\bibitem{pizzo1}
Paul Chleboun, Stefan Grosskinsky, and Andrea Pizzoferrato.
\newblock Lower current large deviations for zero-range processes on a ring.
\newblock {\em Journal of Statistical Physics}, 167(1):64--89, 2017.

\bibitem{chetrite}
Rapha\"{e}l Chetrite and Hugo Touchette.
\newblock Nonequilibrium {M}arkov processes conditioned on large deviations.
\newblock {\em Annales de L'Institut Henri Poincar{\'e}}, 16:2005--2057, 2015.

\bibitem{nyawo}
Pelerine Tsobgni~Nyawo and Hugo Touchette.
\newblock Large deviations of the current for driven periodic diffusions.
\newblock {\em Physical Review E}, 94:032101, 2016.

\bibitem{harris1}
Rosemary~J. Harris and Gunter~M. Schütz.
\newblock Fluctuation theorems for stochastic dynamics.
\newblock {\em Journal of Statistical Mechanics: Theory and Experiment},
  2007(07):P07020, 2007.

\bibitem{hollander}
Frank Den~Hollander.
\newblock {\em Large deviations}, volume~14 of {\em Graduate Texts in
  Mathematics}.
\newblock American Mathematical Society, 2008.

\bibitem{dembo}
Amir Dembo and Ofer Zeitouni.
\newblock {\em Large deviations techniques and applications}, volume~38.
\newblock Springer Science \& Business Media, 2009.

\bibitem{bertini2015}
Lorenzo Bertini, Alessandra Faggionato, and Davide Gabrielli.
\newblock Large deviations of the empirical flow for continuous time {M}arkov
  chains.
\newblock {\em Annales de L'Institut Henri Poincar{\'e}}, 51(3):867--900, 2015.

\bibitem{del2003particle}
Pierre Del~Moral and Laurent Miclo.
\newblock Particle approximations of {L}yapunov exponents connected to
  {S}chr{\"o}dinger operators and {F}eynman-{K}ac semigroups.
\newblock {\em ESAIM: Probability and Statistics}, 7:171--208, 2003.

\bibitem{del2013mean}
Pierre Del~Moral.
\newblock {\em Mean field simulation for {M}onte {C}arlo integration}.
\newblock CRC Press, 2013.

\bibitem{smc:methodology:Bak85}
James E. Baker. 
Adaptive selection methods for genetic algorithms.
{\em In Proceedings of an International Conference on Genetic Algorithms and their applications}, 101--111, 1985.

\bibitem{smc:methodology:KLW94}
Augustine Kong, Jun S. Liu and Wing Hung Wong. Sequential imputations and Bayesian missing data problems. 
{\em Journal of the American statistical association}, 89(425):278--288, 1994.

\bibitem{liggett2010continuous}
Thomas~Milton Liggett.
\newblock {\em Continuous time {M}arkov processes: an introduction}, volume 113
  of {\em Graduate Texts in Mathematics}.
\newblock American Mathematical Society, 2010.

\bibitem{mim}
Stefan Grosskinsky and Watthanan Jatuviriyapornchai.
\newblock Derivation of mean-field equations for stochastic particle systems.
\newblock {\em Stochastic Processes and their Applications}, 2018.
\newblock in press (arXiv:1703.08811).

\bibitem{daipra}
P.~Dai Pra.
\newblock Stochastic mean-field dynamics and applications to life sciences,
  2017.
\newblock http://www.cirm-math.fr/ProgWeebly/Renc1555/CoursDaiPra.pdf.

\bibitem{teicher}
Yuan~Shih Chow and Henry Teicher.
\newblock {\em Probability Theory - Independence, Interchangeability,
  Martingales}.
\newblock Springer, 3rd edition, 1998.

\bibitem{garrahan}
Juan~P Garrahan, Robert~L Jack, Vivien Lecomte, Estelle Pitard, Kristina van
  Duijvendijk, and Frédéric van Wijland.
\newblock First-order dynamical phase transition in models of glasses: an
  approach based on ensembles of histories.
\newblock {\em Journal of Physics A: Mathematical and Theoretical},
  42(7):075007, 2009.

\bibitem{limmer}
Ushnish Ray, Garnet Kin-Lic Chan, and David~T. Limmer.
\newblock Importance sampling large deviations in nonequilibrium steady states.
  i.
\newblock {\em The Journal of Chemical Physics}, 148(12):124120, 2018.

\bibitem{grosskinsky13}
Stefan Grosskinsky, Frank Redig, and Kiamars Vafayi.
\newblock Dynamics of condensation in the symmetric inclusion process.
\newblock {\em Electronic Journal of Probability}, 18(66):1--23, 2013.

\bibitem{bianchi17}
Alessandra Bianchi, Sander Dommers, and Cristian Giardin{\`a}.
\newblock Metastability in the reversible inclusion process.
\newblock {\em Electronic Journal of Probability}, 22(70):1--34, 2017.

\bibitem{cavallaro}
Massimo Cavallaro and Rosemary~J. Harris.
\newblock A framework for the direct evaluation of large deviations in
  non-markovian processes.
\newblock {\em Journal of Physics A: Mathematical and Theoretical},
  49(47):47LT02, 2016.

\end{thebibliography}

\end{document}